\documentclass{emulateapj}

\usepackage{amsmath}

\newcommand{\zptcolor}{\boldsymbol{\kappa}}
\newcommand{\color}{\boldsymbol{c}}
\newcommand{\colormatrix}{\mathbf{B}}
\newcommand{\identity}{\boldsymbol{1}}
\newcommand{\colorairmassmatrix}{\mathbf{T}}
\newcommand{\sdss}{SDSS}
\newcommand{\slr}{SLR} 
\newcommand{\imacs}{IMACS}
\newcommand{\ldss}{LDSS3}
\newcommand{\tmass}{2MASS}

\newcommand{\gof}{GOF} 

\newcommand{\metal}{[\mathrm{Fe}/\mathrm{H}]}

\begin{document}

\title{Stellar Locus Regression: Accurate Color Calibration, and the
  Real-time Determination of Galaxy Cluster Photometric Redshifts }

\author{F.~William High, Christopher W.~Stubbs, Armin Rest, Brian
  Stalder, Peter Challis}

\affil{Department of Physics\\and\\Harvard-Smithsonian Center for
  Astrophysics}

\affil{Harvard University}

\affil{Cambridge, MA}

\email{high@physics.harvard.edu}

\begin{abstract}

  We present Stellar Locus Regression (\slr), a method of directly
  adjusting the instrumental broadband optical colors of stars to
  bring them into accord with a universal stellar color-color locus,
  producing accurately calibrated colors for both stars and galaxies.
  This is achieved without first establishing individual zeropoints
  for each passband, and can be performed in real-time at the
  telescope.  We demonstrate how \slr\ naturally makes one wholesale
  correction for differences in instrumental response, for atmospheric
  transparency, for atmospheric extinction, and for Galactic
  extinction.  We perform an example \slr\ treatment of \sdss\ data
  over a wide range of Galactic dust values and independently recover
  the direction and magnitude of the canonical Galactic reddening
  vector with 14--18 mmag RMS uncertainties.  We then isolate the
  effect of atmospheric extinction, showing that \slr\ accounts for
  this and returns precise colors over a wide of airmass, with 5--14
  mmag RMS residuals.  We demonstrate that \slr-corrected colors are
  sufficiently accurate to allow photometric redshift estimates for
  galaxy clusters (using red sequence galaxies) with an uncertainty
  $\sigma(z)/(1+z) = 0.6\%$ per cluster for redshifts $0.09<z<0.25$.
  Finally, we identify our objects in the \tmass\ all-sky catalog, and
  produce $i$-band zeropoints typically accurate to 18 mmag using only
  \slr.  We offer open-source access to our IDL routines, validated
  and verified for the implementation of this technique, at
  http://stellar-locus-regression.googlecode.com.

\end{abstract} 


\section{Introduction}
\label{sec:intro}

The observed broadband colors---{\it i.e.~}flux ratios---of celestial objects
depend on the photon spectral energy distribution of the source; on
extragalactic, Galactic and atmospheric scattering and absorption
along the line of sight; and on the instrumental sensitivity function
over the wavelengths of interest. One challenge of astronomical
photometric analysis is to disentangle, from a given set of
observations, the source's colors from such a plethora of perturbing
factors.

Colors hold information about a source's type, temperature,
metallicity, and redshift.  A source's apparent magnitude, on the
other hand, also depends on its distance and on the size and nature of
the emitting regions. We assert that for almost all astrophysical
endeavors, accurate photometric {\it colors} are more useful than high
accuracy {\it magnitudes}, especially because we seldom know distances
well enough to convert from apparent to absolute magnitudes at the
percent level.  Furthermore, for photometric redshift techniques that use
a prior on magnitudes, 
 the broad luminosity function of galaxies generates
a span in magnitude that far exceeds the range in color, for a given
galaxy type at a particular redshift.

% What "matters of efficiency"?
% These matters of efficiency are especially important for a timely
% scientific yield by large scale surveys.  

The current and next generation of wide-field multicolor survey
projects include the
CFHTLS\footnote{\url{http://www.cfht.hawaii.edu/Science/CFHLS/}},
PanSTARRS\footnote{\url{http://pan-starrs.ifa.hawaii.edu/public/}},
BCS\footnote{\url{http://cosmology.uiuc.edu/BCS/}},
DES\footnote{\url{https://www.darkenergysurvey.org/}},
LSST\footnote{\url{http://www.lsst.org/Science/lsst\_baseline.shtml}},
and SkyMapper\footnote{\url{http://msowww.anu.edu.au/skymapper/}}.
Object classification and distance estimation with photometric
redshifts are necessary starting points for extragalactic science
using these surveys, and these in turn depend on knowing calibrated
colors. The technique we describe here can be used very early in a
survey to obtain highly accurate colors of objects, as well as
magnitude estimates. This should allow more rapid exploitation of new
survey data---indeed, of nearly any multiband data.

%  Colors are also important for smaller-scale scientific
%programs.
%Our approach has been developed with the current and upcoming
%wide field imagers in mind, but we have used it to good effect with
%instruments with fields of view a few arcminutes across.

The standard approach to determining colors of sources is to first
calibrate magnitudes in all observed passbands (such as $g$, $r$, $i$,
and $z$), and then subtract the calibrated magnitudes to obtain
calibrated colors ($g-r$, $g-i$, {\it etc}).  This is typically
time-consuming, both at the telescope and in the analysis
phase. Establishing photometric zeropoints for a stack of multiband
images requires separate observations of spectrophotometric standard
stars to measure the instrumental sensitivity and estimate atmospheric
extinction, and spatiotemporal interpolation of the calibration
parameters to the science fields under ``photometric conditions.''

In this paper we describe how to calibrate colors directly from
objects cataloged from multiband, flat-fielded images of a field,
without having to first determine the corresponding photometric
zeropoints, and without the usual repeated measurements of standard
stars.  We demonstrate how this can be done accurately, yielding
colors accurate to a few percent, and rapidly, allowing for optimal
use of allocated telescope time.
%Accurate
%colors have numerous applications. 
%One example is star-galaxy
%discrimination, exploiting the universality of the stellar
%locus. Another is the identification of certain classes of stars,
%based on colors alone. 
%Finally, as shown in the example we have presented below, colors can
%be used to select red sequence galaxies and estimate their redshifts.

Our technique exploits the optical and infrared color-color {\it
  stellar locus} \citep[cf.][]{bib:covey,bib:ivezic_stripe82}, the
one-dimensional and astrophysically fundamental track that stars
occupy in color-color space.  The majority of stars lie somewhere
along this locus, at a position that depends primarily upon effective
temperature.  The universality of the stellar locus was exploited in
the Oxford-Dartmouth Thirty-Degree Survey \citep[``stellar locus
fitting,''][]{bib:odts1} to stabilize photometric zeropoints in
non-photometric conditions, and in Sloan Digital Sky Survey's (\sdss)
Stripe 82 \citep[``stellar locus method,''][]{bib:ivezic_stripe82} to
account for differences in the response function of different
detectors in the SDSS instrument.  Our approach is different from
these mainly because we do not first establish photometric zeropoints
per band: we immediately calibrate colors in all fields, and only
optionally solve for the calibrated apparent magnitudes, using the
stellar locus.  We do not use \slr\ as a diagnostic or corrective
tool, but as our primary calibrator.

Our Stellar Locus Regression (\slr) approach builds on these previous
pioneering works.  We establish the location of the stellar feature in
instrumental color-color space, and we determine what transformations
are needed to bring this into coincidence with the known location of a
standard stellar locus.  Applying appropriate color-corrections to the
entire rest of the catalog automatically accounts for all of the
standard calibration terms, including zeropoints, atmospheric
extinction, aperture corrections, and Galactic extinction.  The
technique is straightforward, fast, and allows observers to forgo the
usual standard star observations altogether because all observed
stars are expected to lie along the same stellar locus.  We have calibrated fields
with \slr\ using as few as 7 stars in fields of view as small as
$~4'\times8'$.

\begin{figure}
 \plotone{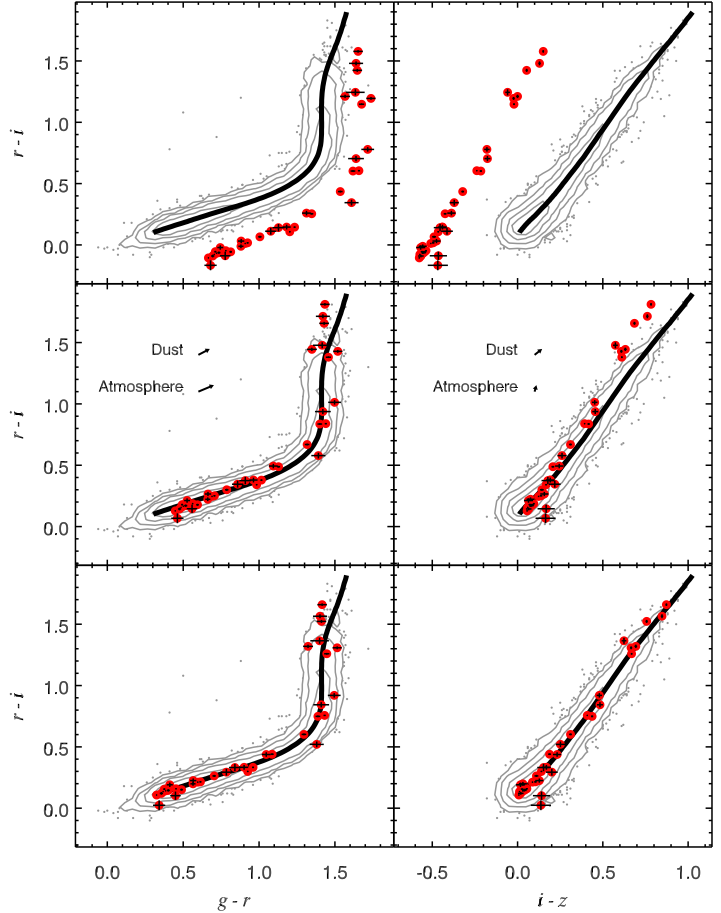}
 \caption{ An illustration of Stellar Locus Regression
   (\slr).  Colors are plotted on the \sdss\ photometric system. All
   panels show the standard stellar locus (black line and gray density
   contours), reproduced from \citet{bib:covey}.  Red points are
   stellar colors obtained from a Source Extractor analysis of
   flat-fielded Magellan $6.5\,\mathrm{m}$ \imacs\ images.  {\it Top
     panels:} The instrumental \imacs\ colors are plotted, with a
   clear mismatch between them and the standard locus. {\it Middle
     panels:} \slr\ is performed with only a common translation vector
   applied to the instrumental colors.  Note the color-dependent
   discrepancies in the upper right portions of the central panels.
   By way of example, the vectors show the expected direction and
   magnitude of extinction by Galactic dust ($A_r=0.2$) and the
   atmosphere ($1.3$ airmasses).  {\it Bottom panels:} Color terms are
   measured from a single observation of a field containing standard
   stars. Fixing these color terms, a new best-fit translation is
   determined, which brings the observed colors onto the
   \sdss-calibrated color system, as defined by the stellar locus.
   This \slr\ analysis, when the corrections are then applied to all
   objects in the photometric catalog, allows us to rapidly obtain
   highly accurate colors on the SDSS system, directly from
   flat-fielded data, with a single correction step that accounts for
   atmospheric extinction, Galactic extinction and instrumental
   response differences.}
 \label{fig:example}
 \end{figure}

Figure \ref{fig:example} schematically illustrates the technique.  
We perform \slr\ on new data (see
\S\ref{sec:colormag}-\ref{sec:photoz}) taken with the \imacs\
instrument \citep{bib:imacs,bib:magellan2008} on the Magellan
$6.5\,\mathrm{m}$ telescope.  All panels show our adopted standard stellar
locus line and stellar density contours, reproduced from
\citet[][\S\ref{sec:standardlocus}]{bib:covey}, along with $36$
\imacs\ stellar colors (red points).  The top panels show instrumental stellar
colors with the standard locus.  We perform \slr, neglecting instrumental color
terms, with results shown in the middle panels.  We perform \slr\
again after measuring color terms independently from an external
standard star field, with results shown in the bottom panels.  By way
of illustration, the vectors in the middle panels show the expected
direction and magnitude of extinction by Galactic dust
\citep[$A_r=0.2\,\mathrm{mag}$ and $R_V=3.1$, estimated using][]{bib:sfd} and the
atmosphere ($1.3$ airmasses).

\epsscale{1.0}

% In this paper we describe the mathematical details and establish a
% useful framework for \slr\ analysis.  We assess how well the \slr\
% approach sidesteps the traditional bottlenecks of atmospheric
% extinction determination and zeropoint calibration. We demonstrate the
% effectiveness of \slr\ in correcting for Galactic and atmospheric
% extinction. An illustrative application is shown in which photometric
% redshifts of galaxy clusters are extracted from \slr-corrected
% colors. We identify and estimate the impact of potential sources of
% systematic error for this technique.

We adopt a standard stellar locus in \S\ref{sec:stellarlocus}, and
then motivate and build a picture of what comprises it so as to
understand its universality.  In \S\ref{sec:method} we outline the
mathematics of color calibration and describe an optimal algorithm for
real-time color calibration, which we apply to real data in a series
of tests (\S\ref{sec:tests}).  In \S\ref{sec:galext} we perform our
first test, applying \slr\ to already-calibrated \sdss\ photometry,
where we recover the canonical Galactic reddening vector in direction
and magnitude over a wide rage of dust thicknesses.  Section
\ref{sec:colormag} examines the fundamental repeatability of \slr\
color and magnitude measurements by comparing \sdss\ data to data from
a different instrument.  Section \ref{sec:atmext} isolates the effect
of atmospheric extinction.  Our final test appears in
\S\ref{sec:photoz}, where we recover the spectroscopic redshifts of
$11$ low redshift galaxy clusters to high accuracy using only \slr\
colors.  We end with a discussion (\S\ref{sec:discussion}) and
conclusions (\S\ref{sec:conclusions}).

%%%%%%%%%%%%%%%%%%%%%%%%%%%%%%%%%%%%%%%%%%%%%%%%%%%%%%%%%%%%%%%%%%%%%%%%%%%%%%%%

\section{The Stellar Locus}
\label{sec:stellarlocus}

The nearly-blackbody emission spectra of stars place them
predominantly along a line in optical and infrared color-color space.
Stellar color therefore depends primarily on effective temperature and
is the basis for the Morgan-Keenan (MK) stellar classification system
\citep{bib:mk}.  Real stellar atmospheres deviate from blackbody
behavior because of molecular absorption and metallicity effects.
Observations, which capture stellar light that has traversed the dust
of our galaxy, our atmosphere, and our telescope systems, will produce
stellar color loci that deviate further from the expected intrinsic
behavior.  We have based our \slr\ approach on the assumption that we
can identify a stellar locus in $grizJHK$ color space that is
intrinsically universal.  We explore the extent to which this is true
by inspecting what comprises an observed stellar locus.

\subsection{A Standard Locus}
\label{sec:standardlocus}

We adopt as our standard the empirical stellar locus of
\citet{bib:covey}.  Those authors calculated the running-median of the
colors of $\sim10^5$ stars, from high quality observations by both the
\sdss\ and \tmass\ surveys.  The line-of-sight Galactic dust for their
sample was estimated from the maps of \citet[][SFD]{bib:sfd} to be
$A_r<0.2$.  We additionally smooth this locus line with a
$0.2\,\mathrm{mag}$ size boxcar averaging kernel to reduce some of the
scatter between adjacent points.  These are the data we present in
Figure \ref{fig:example}---we call this the standard stellar locus
line.

% \begin{figure}
% \plottwo{fig/covey_median_sloan.eps}{fig/covey_median_2mass.eps}
% \caption{ The running-median stellar locus line in $6$-dimensional
%  color-color space from real \sdss\ and \tmass\ data \citep{bib:covey},
%  smoothed here with a 0.2 mag wide boxcar average kernel. }
% \label{fig:locusline}
%  \end{figure}

%Any variation between
%our calibrator locus and the stars in the images being
%color-calibrated with \slr\ will introduce systematic color errors.

%It is remarkable that objects whose absolute luminosities differ by a
%factor of over 1000 (where $(g-r) >1$) still coincide in color-color
%space.

%\subsection{Gross Composition}

The two most salient features of the standard locus are its nearly
one-dimensional nature and a dramatic break or kink in the $(g-r,r-i)$
plane at $r-i\sim0.7$.  As we explore below (\S\ref{sec:colorvolume}),
blue-ward of the kink are mainly evolved and main sequence (MS) A-
through K-type stars, and red-ward are mainly M-type dwarfs
\citep{bib:finlator,bib:sdssMdwarfs,bib:covey,bib:juric}.

Because of the well-defined relationships between the color (effective
temperature), age, and intrinsic luminosity of stars, each point along
the stellar locus probes roughly predictable $3$-dimensional spatial
volumes, given some fixed dynamic range for the observations.  The
reddest MS stars that make up the $r-i>0.7$ branch of the stellar
locus are less luminous, so the effective volume an observation probes
is small and nearby.  Likewise, bluer MS and evolved stars are
intrinsically brighter, so the effective volumes probed at those
colors are larger and farther away---and because both MS and evolved
stars constitute the $r-i<0.7$ portion of the stellar locus,
observations are sensitive to a plurality of volumes and distances.

A comprehensive Galactic structure and population synthesis analysis is
beyond the scope of this paper, but we will explore below the factors that 
give rise to uniformity in the {\it observed} stellar locus, including
the dramatic color-volume effects at typical \sdss\ depths.

\subsection{Color-Volume Effects}
\label{sec:colorvolume}

Each pointing of an astronomical camera images a cone-shaped region of
space, set by the solid-angle field of view $\Omega$ of the
instrument. For a single frame there is also a limited observable
dynamic range in apparent magnitude: the brightest observable objects
are determined by the saturation limit of the system, and the faintest
useful objects must satisfy some selection in signal-to-noise ratio.
A star of a particular absolute magnitude and corresponding color is
therefore detectable within a truncated cone of opening angle
$\Omega$, with an near-edge determined by the saturation limit and a
far-edge set by the required signal-to-noise ratio.  For simplicity we
assume an unextincted line of sight.

Table \ref{tab:volume} illustrates the dramatic selection-effect of
dynamic range on observed stars of various intrinsic luminosities.  We
take a notional saturation limit of $r=14$ and a faint detection limit
at $r=22$, which are approximate values appropriate for \sdss.  For a
dynamic range of $8\,\mathrm{mag}$, the outer detection edge of the
detectability cone is always $\sim39$ times farther away than its
inner saturation edge, for a given absolute magnitude object.  Table
\ref{tab:volume} also shows that for each additional magnitude
increase of stellar luminosity, the survey volume increases by a factor of four.
Luminous stars are detectable over a vastly larger volume than fainter
stars, since the outer edge of the detectability region is {\it
  proportional to} the distance to the inner edge.

As a concrete example, at \sdss\ depths and Galactic latitudes
$|b|=90\,\deg$, absolute magnitudes $\mathrm{M}_r\gtrsim12$ are
detectable only closer than $\sim1\,\mathrm{kpc}$, which is the height
of the Galactic disk \citep{bib:juric}.  Objects with
$\mathrm{M}_r\lesssim4$ are seen only at Galactic heights
$|Z|>1\,\mathrm{kpc}$, so are in the halo.

\begin{deluxetable}{cccc}
\tabletypesize{\small} % 11pt
%\tabletypesize{\footnotesize} % 10pt
%\tabletypesize{\scriptsize} % 8pt
%\rotate
  \tablewidth{0pt} 

  \tablecaption{Detectability Volumes vs.\ Absolute
    Magnitude.\label{tab:volume} }

  \tablehead{ 
    \colhead{M$_r$\tablenotemark{a}} & 
    \colhead{$R_{\mathrm{inner}}$\tablenotemark{b}} &
    \colhead{$R_{\mathrm{outer}}$\tablenotemark{c}}  & 
    \colhead{Detectability Volume\tablenotemark{d}} \\
    \colhead{(AB mag)} & \colhead{(pc)} & 
        \colhead{(pc)} & \colhead{(arb units)} }

\startdata
-5 & 6.3E4 & 2.5E6 & 1.0E12 \\
-4 & 4.0E4 & 1.6E6 & 2.8E11 \\
-3 & 2.5E4 & 1.0E6 & 6.3E10 \\
-2 & 1.6E4 & 6.3E5 & 1.6E10 \\
-1 & 1.0e4 & 4.0E5 & 4.0E9 \\
0 & 6.3E3 & 2.5E5 & 1.0E9 \\
1 & 4.0E3 & 1.6E5 & 2.5E8 \\
2 & 2.5E3 & 1.0E5 & 6.3E7 \\
3 & 1.6E3 & 6.3E4 & 1.6E7 \\
4 & 1.0E3 & 4.0E4 & 4.0E6 \\
5 & 6.3E2 & 2.5E4 & 1.0E6 \\
6 & 4.0E2 & 1.6E4 & 2.5E5 \\
7 & 2.5E2 & 1.0E4 & 6.3E4 \\
8 & 1.6E2 & 6.3E3 & 1.6E4 \\
9 & 1.0E2 & 4.0E3 & 4.0E3 \\
10 & 6.3E1 & 2.5E3 & 1.0E3 \\
11 & 4.0E1 & 1.6E3 & 2.5E2 \\
12 & 2.5E1 & 1.0E3 & 6.3E1 \\
13 & 1.6E1 & 6.3E2 & 1.6E1 \\
14 & 1.0E1 & 4.0E2 & 4.0E0 \\
15 & 6.3E0 & 2.5E2 & 1.0E0 \\

\enddata

\tablenotetext{a}{Absolute magnitudes of stars in the \sdss\
  $r$-band.}

\tablenotetext{b}{Closest distances (in parsecs) for the object to be
  observable, for a notional saturation limit of $r=14$.}

\tablenotetext{c}{Furthest distances (in parsecs) for a $10\sigma$
  detection limit of $r=22$.}

\tablenotetext{d}{Volume (in arbitrary units) within which objects of
  a given absolute magnitude can be detected, subject to the
  instrumental dynamic range constraints.}

\end{deluxetable}

%To arrive
%at the relative number of stars probed, these volume figures must be
%taken together with stellar number density expectation.  For example,
%looking toward Galactic latitudes of $|b|=90\deg$ and taking a disk
%stellar number density that scales as $\rho_{\mathrm{disk}}(z)\sim
%\rho_0\,e^{-z/z_0}$ with $z_0=1\,\mathrm{kpc}$, the peak in star
%counts occurs at a vertical distance of $2\,\mathrm{kpc}$, due to the
%monitored volume increasing as $z^2 \mathrm{d}z$.

%\subsubsection{Stellar Populations in Color-Color Space}

\begin{figure}
  \plotone{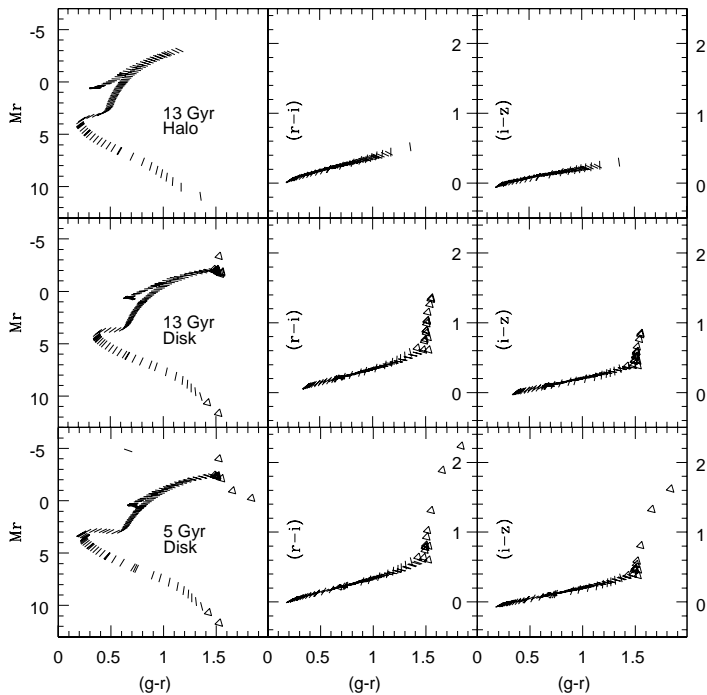}
  \caption{Comparisons of modeled $M_r$ vs.\ $g-r$ color-magnitude
    diagrams (left column), $r-i$ vs.\ $g-r$ (central column), and
    $i-z$ vs.\ $g-r$ (rightmost column) color-color diagrams for
    different stellar populations, for zero Galactic extinction.  The
    upper row corresponds to a stellar halo population with
    $\metal=-1.5$ and an age of 13 Gyrs.  The central row also
    corresponds to a 13 Gyr age, but with a characteristic disk
    metallicity; $\metal=-0.5$. The bottom row is what one might
    expect for more recent star formation in the Galactic disk, with
    $\metal=-0.5$ and an age of 5 Gyr.  Absolute magnitudes are
    indicated by the {\it angle} of the whisker, with
    $\theta=9M_r\,\mathrm{mag}^{-1}\,\deg$. Objects with $r-i>0.7$ are
    shown as triangles.  Only stellar populations with $\metal> -0.7$
    produce the excursion to $r-i>0.7$, seen in the central and lower
    rows, that is essential to the \slr\ method.}
 \label{fig:comparison}
\end{figure}

Absolute stellar magnitude varies with stellar color and age in well
understood ways.  To explore the relation in the \sdss\ magnitude
system, we used the model stellar populations of
\citet[][\url{http://stev.oapd.inaf.it/cgi-bin/cmd\_2.1}]{bib:isochrones}.
This online tool modeled stars with initial masses down to
$0.15\,M_\sun$.  We have just shown that the typical \sdss\ dynamic
range is sensitive to both halo and disk stars, so we consider three
models, using \citet{bib:ivezic_metallicity} as a guide:
\begin{enumerate}
\item a $13\,\mathrm{Gyr}$  halo population, with a metallicity
  $\metal_{\mathrm{halo}}=-1.5$,
\item a $13\,\mathrm{Gyr}$ disk population of higher metallicity,
  $\metal_{\mathrm{disk}}=-0.5$, and
\item a $5\,\mathrm{Gyr}$ disk population of higher metallicity,
  $\metal_{\mathrm{disk}}=-0.5$.
\end{enumerate}
Figure \ref{fig:comparison} shows the color-magnitude and color-color
diagrams that illustrate the connection between absolute magnitudes
and regions in color-color space for these populations.  

Figure \ref{fig:comparison} shows that only disk-metallicity stars 
populate the locus red-ward of the kink.  The main sequence M dwarfs
in this color range have $M_r\gtrsim11\,\mathrm{mag}$ and so in the \sdss\ 
dynamic range these objects lie at distances
$\lesssim1.5\,\mathrm{kpc}$ from Earth.  These stars are almost entirely
in the disk.  The M giants red-ward of the kink have
$M_r\lesssim-2\,\mathrm{mag}$ corresponding to distances $\gtrsim15\,\mathrm{kpc}$.  
Few disk-metallicity stars lie at such
distances \citep{bib:ivezic_metallicity}.  Therefore the vast majority of
the stars that populate the $(r-i)>0.7$ region of the locus at \sdss\ depths are disk
dwarfs.  \citet{bib:sdssMdwarfs} also support this viewpoint,
attributing \sdss\ objects with $1<(r-i)<2$ to M0 through M6 stars at
typical distances ranging from $1$ to $0.1\,\mathrm{kpc}$ respectively.

Continuing these arguments, stellar locus stars blue-ward of the
$(r-i)\sim0.7$ kink comprise of a blend of MS disk stars, MS halo stars, and 
evolved (post-turnoff) halo stars.  The 
luminosity of the bluer MS stars are $4\,\mathrm{mag}\lesssim
M_r\lesssim11\,\mathrm{mag}$ and are detectable from both the disk and halo.  We also
see that even for a stellar population $13\,\mathrm{Gyrs}$ old, with a
main sequence turnoff at around $M_r\sim4$, blue turnoff stars that
are closer than $1\,\mathrm{kpc}$ would saturate the detector. 

The relative proportion of disk and halo stars that make up the stellar
locus at $(r-i)<0.7$ (blue-ward of the kink) is magnitude dependent. 
\citet{bib:ivezic_metallicity} show that for $0.2<(g-r)<0.4$,  at $g\sim17.5$  
half the stars have metallicity characteristic of the disk, and half are halo stars. 
At fainter magnitudes the proportion of halo stars increases. 

Armed with this understanding of the observed stellar locus's constituent
populations, we explore the effect their corresponding metallicity
distributions should have.

\subsection{Metallicity}

A number of previous authors have explored the influence of
metallicity on stellar colors in the \sdss\ photometric system
\citep{bib:properties,bib:covey,bib:juric,bib:ivezic_metallicity}.
\citet{bib:ivezic_metallicity} demonstrated that most halo stars have
a common metallicity that is reasonably well described by a Gaussian
distribution with median $[\mathrm{Fe}/\mathrm{H}]_{\mathrm{halo}}=-1.5$ and
width $\sigma_\mathrm{halo}=0.3\,\mathrm{dex}$.  The median disk
metallicity was shown to be about $-0.5$ and exhibits half the scatter
of the halo, with $\sigma_\mathrm{disk}=0.16\,\mathrm{dex}$.

The biggest observable broadband effect from metallicity occurs in the
$u$-band, which is one of the reasons we exclude that band from
consideration here.  Stellar metallicity effects in $g-r$ are ten
times smaller than in $u-g$.  Using this fact and Figure 21 of
\citet{bib:ivezic_metallicity}, the expected spread in $g-i$ color for
halo stars should be of order $0.05\,\mathrm{mag}$.  This is a
conservative upper limit estimate of the intrinsic width of the halo
contribution to the stellar locus from metallicity alone.

We have modeled the effect of metallicity on the stellar MS, and
present some of the results in Table \ref{tab:metallicity}.  We used a
grid of model atmospheres derived from the Phoenix project
\citep{bib:phoenix} and the \citet{bib:kurucz} library for the typical
disk and halo metallicities, $\metal=-0.5$ and $-1.5\,\mathrm{dex}$.
The range of parameters used was $3,500K \le T_{\mathrm{eff}} \le
10,000K$ for M5 to A0 using the Phoenix models and $15,000K \le
T_{\mathrm{eff}} \le 45,000K$ for B5 to O5 from the \citet{bib:kurucz}
models, and we adopted a typical main sequence $\log(g)$ for each
spectral type.  We then calculated the broadband colors based on the
SDSS filter and CCD throughput.

\begin{deluxetable}{cccc}
%\tabletypesize{\small} % 11pt
\tabletypesize{\footnotesize} % 10pt
%\tabletypesize{\scriptsize} % 8pt
%\rotate
  \tablewidth{0pt} 
  \scriptsize

  \tablecaption{Model Stellar Color Difference Between Z=$-$0.5 and
    Z=$-$1.5.\label{tab:metallicity} }

  \tablehead{ 
    \colhead{MK type}&%\tablenotemark{a}} & 
    \colhead{$\delta(g-r)$\tablenotemark{a}} &
    \colhead{$\delta(r-i)$\tablenotemark{a}} &
    \colhead{$\delta(i-z)$\tablenotemark{a}}
}

\startdata
%\cutinhead{Change from Z=$-$0.5 to Z=$-$1.5}
O5V & -0.005 & -0.002 & -0.003 \\
B0V & -0.000 & -0.002 & -0.002 \\
B5V & -0.003 & -0.002 & -0.005 \\
A0V & -0.000 & -0.004 & -0.003 \\
A5V & 0.004 & -0.007 & -0.007 \\
F0V & 0.004 & -0.010 & -0.012 \\
F5V & 0.019 & -0.009 & -0.013 \\
G0V & 0.031 & -0.012 & -0.015 \\
G5V & 0.037 & -0.013 & -0.015 \\
K0V & 0.053 & -0.015 & -0.014 \\
K5V & 0.124 & 0.002 & -0.014 \\
M0V & 0.004 & 0.067 & 0.031 \\
M5V & -0.137 & 0.117 & 0.080 \\
\enddata
\tablenotetext{a}{As a comparison, for galactic extinction ($A_r=0.2$) 
the color perturbation is (0.076, 0.048, 0.044); for 1.3 atmospheres of 
extinction, the color perturbation is (0.103, 0.053, 0.010).}  
\normalsize
\end{deluxetable}

We found that the greatest influence of metallicity on broadband color
is in the late-type stars (around M0 and later) and is in opposite
direction to the color displacement of early-type stars.  
The relatively cool atmospheres of the red stars
are rich with molecules, and the molecular absorption features blend
together with other lines to form a continuum that differs
substantially from a blackbody spectrum.  We also observed that a
minimum metallicity of $[\mathrm{Fe}/\mathrm{H}]=-0.7$ is needed for red
stars
% in fairly mature stellar populations (age $>5\,\mathrm{Gyr}$)
to populate the locus red-ward of the kink.  Less metallicity
``irons out'' the kink, making the model colors colinear with the
bluer stars.  This observation is consistent with the conclusions
drawn in \S\ref{sec:colorvolume}.  A continuum of metallicities near
this threshold would smear the observed stellar locus in the redder
region. The apparent lack of such smearing in the observed locus
indicates the paucity of stars at this intermediate metallicity, as
found by \citet{bib:ivezic_metallicity}.  It is therefore the high
mean metallicity and molecular absorption of disk stars that produce the kink in the \sdss\
stellar locus in the first place, and that make our implementation of
\slr\ possible.

\subsection{Age}

Younger stellar populations have a brighter, bluer main sequence
turnoff point, and this shifts the blue tip of the stellar locus
towards the blue.  This does not distort the basic shape of the
stellar locus.  For very young populations at high metallicity 
there is also a red-ward extension of the main locus,
due to massive stars evolving to $r-i>0.7$, but we concluded in
\S\ref{sec:colorvolume} that these stars do not appear in the stellar
locus at \sdss\ depths, since these young stars would be in the disk and 
would saturate.  Other features in the color-magnitude
diagram, such as helium-burning loops, essentially remain confined to
the standard locus line and do not change its shape.  Our methods
(\S\ref{sec:method}) are therefore expected to be largely insensitive to population
age effects.

%FWH: As per Chris's request a while back, I truncated the blue tip because it appeared noisy.  Covey's data goes bluer than this.
% The fiducial locus we adopt (\S\ref{sec:standardlocus}) has a blue tip
% that is truncated at about $g-r\sim0.2$.
% On the other hand it also
% portrays an extension to $r-i>1.5$ that seems indicative of the more
% massive, evolved component of a younger population.

\subsection{Binaries and Unresolved Superpositions}

Superpositions of stars, either as physical binaries or chance
coincidences, most often fall within the standard locus area, but can
also produce systematic outliers and therefore change the observed
stellar locus shape.  For the latter to happen, the objects within a
single PSF must have different colors, but similar magnitudes.  One
can adopt two philosophical resolutions to this issue.

One approach is to realize that both our standard locus and the
instrumental colors should contain binaries, and so the effect will
average to zero over many fields.  While this may be true
statistically, if the stars drawn from an observed field are few in
number, the observed stellar locus may be distorted by small number
statistics.

Another approach is to iteratively excise stellar colors that are
significantly different from the standard color expectation.  This
should suppress locus shape distortions due to the occasional
unresolved binary, but nevertheless the incidence of binary outliers
is expected to be low.  \citet{bib:binaries} have shown the existence
of an ``echo'' of the stellar locus, which they ascribed to unresolved
binaries of similar luminosity but different effective
temperatures. They assessed the number of objects in this ``echo
locus'' as being fewer than 1/2000 as numerous as the objects that
occupy the main stellar locus.  The combined low incidence and known
location of these outliers makes these binaries unproblematic.
%Regarding possible unresolved binary contamination of the
%sample being subjected to \slr\ analysis, only if we have over 2000
%stars would we expect, on average, a single binary color contaminant.
%This would be rejected by sigma-clipping anyway, if it was a locus
%outlier.

\subsection{Variable Stars}

Photometric variability is a source of systematic error for stellar
locus color methods.  If a star's observed brightness varies in the
time between observations in the various passbands under analysis,
even if the underlying stellar color does not vary, the magnitude
differences will introduce a shift in color-color space. Since the
\sdss\ photometric data in different bands are obtained within a few
minutes of each other, the \sdss-band standard stellar locus should
have minimal contamination from this effect. Similarly, the \tmass\
instrument obtained $J$, $H$, and $K$ photometry essentially
simultaneously.  Combining \tmass\ and \sdss\ photometry, however, can
fall prey to variability, since the observations were taken at
different times. So establishing the joint optical/NIR standard
stellar locus must attend to this issue.

Similarly, if the images under analysis are obtained at different
times, variability on this timescale will distort the derived
calibration color shifts. \citet{bib:sdss_variables} explored the
incidence of variables in the SDSS color-color diagram. They showed
that there is a substantial variation in the fraction of objects that
exhibit variability, across the color-color diagram. The blue tip of
the stellar locus has two classes of variables: low redshift QSOs and
RR Lyrae stars. These can be suppressed by selecting a judicious
region in color-color space where the standard locus is matched to the
instrumental color distribution, iteratively.

The multi-epoch analysis of \sdss\ photometry in
\citet{bib:sdss_variables} indicates that fewer than $5\%$ of the
objects in the main stellar locus exhibit variability with $\Delta g >
0.05\,\mathrm{mag}$. A sigma-clipped iterative analysis of multi-epoch
photometry should be able to produce a cleansed standard stellar
locus.

%(Note to us: temporal variability is a good way to tell red giants from 
%main sequence M stars. Could select a population of halo giants by looking 
%for red, variable stars!)

\subsection{Galactic Dust}
\label{sec:dust}

Galactic dust is another extrasolar source of stellar locus
perturbations.  The dust is shown by \citet{bib:2mass_extinction} and
\citet{bib:juric} to be confined to a sheet roughly $100\,\mathrm{pc}$
above and below the Galactic midplane.  At \sdss\ depths, the vast
majority of stars are therefore behind the dust.

To first approximation the extinction obeys the canonical $R_V=3.1$
reddening law and the degree of extinction $A_V$ follows the maps of
SFD.  This induces a simple overall color-color vector
shift---reddening---whose direction depends on $R_V$ and whose
magnitude depends on $A_V$.  \citet{bib:ivezic_stripe82} showed that
the stellar locus position reflects these predictions in some regions
of \sdss\ Stripe 82, but breaks down as Galactic latitude $|b|$
decreases (\S\ref{sec:galext}).

The usual adoption of extinction coefficients from the SFD appendix
assumes an underlying spectral energy distribution typical of an
elliptical galaxy.  This is not valid for Galactic stars
\citep[eg][]{bib:mccall}.  To next approximation, a more correct
treatment would provide extinction coefficients for each stellar
spectral type.  Errors of this form induce differential distortions of
the stellar locus as a function of effective temperature.

Other deviations occur when the basic assumptions of extinction
behavior break down.  If the reddening law is not strictly $R_V=3.1$
\citep[eg][]{bib:larson}, the stellar locus shifts in different
directions.  If some observed stars are in front of the dust and
others behind, the stellar locus will show additional scatter,
preferentially in the reddening directions.  Because the stars are
most often behind the dust, Galactic extinction is a major contributor
to the observed stellar locus properties.

The \slr\ approach makes no
specific assumption about any value of $R_V$. It corrects for 
Galactic extinction subject to the assumption that the
observed stars all lie behind a {\it common} Galactic extinction layer.
We suspect a fruitful approach might be to make an initial adjustment 
to the instrumental colors using SFD and $R_V=3.1$, and then run 
\slr\ on the resulting catalog. 

\subsection{Summary}

Taking all this information together, we have clear expectations for
what constitutes the typical observed stellar locus, and what factors change
its observed position or shape.

Stars with $r-i>0.7$ are mainly faint M dwarfs in the disk, with
correspondingly high metallicity. Since these objects are only visible
out to $\sim 1\,\mathrm{kpc}$, only metallicity variations in this local
region of the Galaxy could perturb the kink region of the observed
locus.

The stars blue-ward of the kink, however, are a magnitude-dependent
combination of halo stars and disk main sequence stars. The metallicity
dependence of this region of the locus is small.

While the location of the bluest terminating edge of the stellar locus is
an indicator of the age of the underlying population, it does not
distort the basic shape of the locus line. 

Galactic dust should be a prime source for locus shifts and possibly
small distortions.  At high Galactic latitudes the vast majority of
\sdss\ stars reside behind the dust lanes of the Milky Way.  At low
Galactic latitudes, even distant stars come closer to the Galactic
plane.  We would therefore expect to eventually see a smearing of the
stellar locus due to stars suffering different amounts of extinction
along the line of sight.

Finally, atmospheric and instrumental effects will naturally perturb
the stellar locus, and we model these in our mathematical
formulation of Stellar Locus Regression (\S\ref{sec:method}).  Given
what we have learned about the nature of the observed stellar locus,
our conception of \slr\ begins with a simple set of assumptions.
\begin{enumerate}
\item The standard stellar locus is representative of the typical
  stellar populations that we will observe in practice, and is sufficiently 
  uniform so as to constitute a calibration standard. The standard locus stars:
\begin{enumerate}
\item are disk dwarfs in the red and both disk and halo stars in the blue, thanks to
  the dramatic color-volume effects at \sdss\ depths, and
\item lie at high enough Galactic latitude to put them behind the
  dust, but suffer low dust extinction.
\end{enumerate}
%\item The standard locus is effectively calibrated to {\it in front
 %   of} the dust sheet and to the top of Earth's atmosphere.
\item Stars we observe in practice are always {\it behind} the
  Galactic dust sheet.  This way, \slr\ directly outputs dereddened
  colors.
\item The Galactic dust extinction can obey {\it any} power law $R_V$.
\item The Galactic dust extinction is locally smooth.
\item Observations include the $g$-band, and are deep enough to
  observe at least a few of the faint M dwarfs that constitute the
  kink feature \citep[limiting
  $g\gtrsim18\,\mathrm{mag}$,][]{bib:ivezic_metallicity}.
\item The input images from which instrumental photometry is extracted
  are properly flat-fielded.
\end{enumerate}
Many of these are testable in isolation (\S\ref{sec:tests}), but some
possibly degenerate effects, such as simultaneously anomalous
metallicities and extinction laws, may be difficult to disentangle.
We deduce nonetheless that the combination of magnitude dynamic range
selection effects and the relative insensitivity of the $grizJHK$
stellar locus to stellar metallicity, age differences, and binary and
variable contamination produces an observed stellar locus that is
uniform enough to achieve calibration of colors using a standard
stellar locus.

%%%%%%%%%%%%%%%%%%%%%%%%%%%%%%%%%%%%%%%%%%%%%%%%%%%%%%%%%%%%%%%%%%%%%%%%%%%%%%%%

\section{The Method}
\label{sec:method}

The basis of Stellar Locus Regression is to transform instrumental stellar
colors so that they align with a standard locus, on the \sdss\ 
photometric system.  This requires
typical data preprocessing and photometry that produces instrumental stellar colors
from single-epoch, flat-fielded images.  The only calibration images 
required for \slr\ are a single set of multiband observations of
a high-density standard star field, from which instrumental color terms
are measured.  Periodic updates to the instrumental color terms will depend on the 
timescale over which these terms evolve. Mosaic 
imagers will benefit from chip-by-chip color terms. 
Then, with fixed color terms applied,
\slr\ calibrations are performed on science frames by iteratively
transforming the instrumental stellar colors to optimize a goodness-of-fit
(\gof) statistic.  The resulting best-fit parameters, including the instrumental 
color terms, define the color transformation that achieves the \slr\ calibration.
Uncertainties in the calibration are estimated numerically.  These calibration
terms are then applied to all the cataloged photometry of objects appearing
in the same images.  This way all objects in the field are calibrated
using the same stars lying in that field, and no spatial nor temporal
interpolation is required.

In this section we describe and motivate our calibration equations,
discuss our chosen \gof\ and method of error estimation, and outline a
practical algorithm that we have implemented for the real-time
calibration of $griz$ colors.

\subsection{Color Transformations}
\label{sec:transform}

A {\it color transformation} is the mathematical transformation of
colors by translations, scalings, rotations and shears.  Instrumental
colors are represented with the vector\footnote{Our convention is to
  assign vectors boldface, lowercase letters, and matrices boldface,
  capital letters.  The {\it elements} of vectors, matrices, and
  tensors, like all scalars, are not boldface.}  $\color$, and
``true''  colors, which we take to be on the \sdss\ photometric system, 
 are represented with the vector
$\color_0$.  Instrumental colors relate to the true colors via the
color transformation equation
\begin{equation}
  \label{eqn:transform}
  \color = \zptcolor + (\identity + \colormatrix)\color_0.
\end{equation}
The color translation vector $\zptcolor$ accounts for first order
atmospheric extinction, Galactic extinction, zeropoints, and any other
additive effects, {\it known or unknown}.  We discuss $\zptcolor$ in
\S\ref{sec:kappa}.  The color term matrix $\colormatrix$ is populated
by zeros and constants, typically small, and corresponds to the 
instrumental color
terms from conventional photometric calibrations.  We discuss
$\colormatrix$ in \S\ref{sec:colormatrix}.  Appendix \ref{app:corresp}
shows in mathematical detail how Equation \eqref{eqn:transform}
relates to photometric calibration equations that astronomers
typically adopt.

\subsection{The Color Translation Vector}
\label{sec:kappa}

The color translation vector $\zptcolor$ accounts for differences of
zeropoints $a$, atmospheric extinction $E$, and Galactic extinction
$A$.  The scalar elements of the translation vector are
\begin{equation}
  \label{eqn:kappa}
  \kappa_{nm} = a_n - a_m + E_n - E_m + A_n - A_m.
\end{equation}
where $n$ and $m$ are elements of the filter set, for example
$n,m\in\{g,r,i,z\}$, and $n\neq m$.  See Appendix \ref{app:corresp}
for a derivation of this relation, starting from traditional
photometric calibration equations.

In traditional photometric calibration, the terms on the right-hand
side of Equation \eqref{eqn:kappa} are estimated independently.
Atmospheric extinction is usually modeled using a Bouger extinction
law, $k_n X_n$, where $k_n$ is a filter-dependent constant and $X_n$
is the airmass through which the $R$-band image was taken.  The
atmospheric extinction constants are extracted using intermittent
standard star observations at a range of airmasses and interpolated in
space and time to the science frames.  This assumes that the linear
airmass model reflects truth and is spatiotemporally invariant.
Galactic extinction is normally estimated from SFD---a procedure that
assumes a single $R_V=3.1$ dust reddening law \citep[cf.][]{bib:sfd}.
Finally, zeropoints are estimated using the same standard star frames
used to measure the atmospheric terms.  The zeropoints standardize the
photometry, and account for differences in
instrumental throughput between facilities and any other unmodeled
additive effects.  Zeropoints are then interpolated in space and time
to the science frames in ``photometric conditions,'' which again assumes
spatiotemporal invariance.

\slr\ is fundamentally different and is not subject to these
assumptions.  \slr\ fits for each element of the vector $\zptcolor$
directly, and therefore calibrates the entire right-hand side of
Equation \eqref{eqn:kappa} in one step.  The atmospheric and dust
extinctions and the zeropoints are not estimated independently.  This
is made possible by the universality of colors of abundant MS stars
(\S\ref{sec:stellarlocus}).  Because of this unique ability of \slr,
{\it it does not matter what mathematical form the extinction and
  zeropoint terms take in reality, as long as they are
  additive, because all additive systematic effects are accounted for
  together by $\zptcolor$ during \slr.}  For example, the Bouger
atmospheric extinction law is likely incorrect in the $z$-band, where
water absorption is saturated and the additive airmass dependence
probably better follows an $X_z^{1/2}$ airmass power law
\citep[cf.][]{bib:atmos}.  Further, as we have mentioned it is
probable that not all Galactic dust columns obey the canonical
$R_V=3.1$ extinction law.  \slr\ is immune to additive mis-modeling of
the atmosphere, Galactic dust, overall instrument sensitivity
differences, aperture corrections, and so on.

Figure \ref{fig:example} demonstrates the effect of leaving the color
translation vector free during Stellar Locus Regression.  The top
panels show how the known, standard locus compares to an instrumental stellar
locus.  In the middle panel, we perform \slr, letting $\zptcolor$ be
free and fixing the color term matrix $\colormatrix$ to zero.
Residual systematics are clearly evident, in the form of deviations
that appear to increase with color.  This is expected because we are
comparing standard \sdss\ data to data taken with a different
instrument than \sdss.  We remedy this by measuring color terms for
the instrument once using a standard stellar field observation, and
then fixing these nonzero color terms while performing \slr.  Color
term issues are described in the following section.

\subsection{The Color Term Matrix}
\label{sec:colormatrix}

The entries of the color term matrix $\colormatrix$ are zeroes or
constants.  For example, for the \sdss\ colors we use in
\S\ref{sec:colormag}-\ref{sec:photoz}, $\color=(g-r,r-i,i-z)$, we adopt
the color term matrix
\begin{equation}
  \colormatrix =
\begin{pmatrix}
b_g & -b_r   & 0         \\
0     & b_r  & -b_i      \\
0     & 0      & b_i-b_z \\
\end{pmatrix}.
\end{equation}
This has a direct correspondence with traditional photometric color
term formulations, as shown in Appendix \ref{app:corresp}.  

Color terms, and the color term matrix, account for broad,
differential instrument sensitivity differences that arise when data
are acquired with different telescopes, CCDs, or filters than those
used to generate the standard catalog.  For example, we see a clear
systematic error between \imacs\ and \sdss\ in Figure
\ref{fig:example} that varies monotonically with color.  By estimating
color terms with an observation of a standard star field and applying
the color term matrix transformation during our \slr, we significantly
improve the fit.

We have also estimated color terms directly with \slr\ by letting the
color terms be free in addition to $\zptcolor$ during the fit.  This
results in a divergent regression when using our weighted color
residual \gof\ (\S\ref{sec:gof}).  The reason is that the global
minimum of this \gof\ occurs when all instrumental data points
collapse to a single point, which is allowed by divergent shears and
rescalings from $\colormatrix$, and divergent, compensatory color
translations that put the singular instrumental data somewhere on the
standard locus line.  The best-fit color terms are extremely large,
violating our assumption of smallness.  We have experimented with
other GOFs with varying success, but none yields color terms with
accuracy that rivals that of the traditional procedure.

This is the only step in our real-time color calibration procedure
(\S\ref{sec:algorithm}) that requires standard star observations.  If
instrumental color terms are stable over, say, month- and year-long
timescales, then color terms need only be estimated as infrequently.
Without the requirement of multiple standard field exposures per
night, the observer maximizes the total exposure time on science
fields.

\subsection{Color-Airmass and Higher Order Corrections}

Color-airmass terms take the same essential form as instrumental color
terms.  Appendix \ref{app:corresp} gives an explicit example of this.
Correspondingly, color-airmass terms can be estimated with the same
color term procedure described in \S\ref{sec:colormatrix} and
\S\ref{sec:algorithm}.  We have not yet implemented this as we assume
these corrections to be small; this is an obvious future addition to
our \slr\ formulation.  In principle, corrections proportional to
higher order powers of the color, airmass, and even the Galactic dust
column can also be measured, but these will require larger, dedicated
programs in order to minimize error.  As shown in
\S\ref{sec:colormag}-\ref{sec:photoz}, we achieve $1$--$2\%$ level
self-consistency with respect to zeropoints, airmass, and Galactic
dust, without making these corrections.

We now show how, using the same formalism and the same standard locus
of \S\ref{sec:standardlocus}, \slr\ can be made to output individual
calibrations for each filter, producing calibrated photometry instead
of colors.

\subsection{Photometric Calibration with \slr\ Using \tmass}
\label{sec:2mass}

If the stellar locus is extended into other passbands that are already
photometrically calibrated, then the instrumental photometry can be
directly calibrated using only the \slr\ methodology.  \tmass\ is an
obvious choice for an external catalog, as it is full-sky and freely
available and the standard locus (\S\ref{sec:standardlocus}) bridges
$JHK$ with $ugriz$.

The procedure requires the additional prior step of cross-correlating
instrumental stellar catalogs against \tmass's calibrated data.  This is easily
done using the Gator web interface\footnote{\url{http://irsa.ipac.caltech.edu/cgi-bin/Gator/nph-dd?catalog=fp\_psc}}.

For illustration, say we have obtained the $J$-band photometry of our
instrumental $griz$ stars.  We construct a color that is a hybrid of
instrumental optical magnitudes and calibrated \tmass\ magnitudes,
$z-J$, which extends the instrumental stellar locus to
$\color=(g-r,r-i,i-z,z-J)$.  \slr\ is executed in precisely the same
way as for the optical data alone, using the standard optical-infrared
hybrid locus line.  The result is still $\zptcolor$, but the last
entry is $\kappa_{zJ}=a_z+E_z+A_z$ because the \tmass\ data are
already calibrated: $\kappa_{zJ}$ is the $z$-band photometric
calibration.  This can be done for any combination of instrumental
calibrated \tmass\ data and instrumental optical or infrared data from
a different passband.  Appendix \ref{app:corresp} makes the
mathematics of \slr\ photometric calibration explicit for this
particular example.

\slr\ photometric calibration is subject to the errors of \tmass, and
is more precise when more stars are used in the fit.  Given typical
\tmass\ errors of $5\%$, \slr-calibrated photometric zeropoints should
typically be accurate to $5\%$ or better.  By the nature of \slr, the
Galactic dust correction is inextricable from the photometric
calibration when the instrumental stars are all behind the dust, so uncertainty
from the dust must be added in quadrature.  Finally, color terms carry
with them additional uncertainty from the color term estimate and the
colors that they multiply.  If color terms are used, then the colors
must be estimated independently from the the non-\tmass\ data, either
simultaneously during the \tmass\ fit or prior to it.

\tmass\ matches will probably only occur for a {\it subset} of one's
instrumental data.  The reduced number of stars will generically degrade the
Stellar Locus Regression errors.  Our optimized procedure is to first
perform an instrumental-only \slr\ to estimate the colors using a maximal number
of stars, and, if the photometry is needed at all, to perform a
\tmass\ \slr\ on the subset of matches to calibrate the photometry
separately.  This ensures that the color errors are minimal and not subject
to the errors from \tmass\ and from the reduced statistics.  \slr\
never requires the subtraction of calibrated photometry to arrive at
colors.

\subsection{Goodness of Fit}
\label{sec:gof}

During Stellar Locus Regression we optimize goodness-of-fit (GOF)
statistics and estimate errors numerically.  The GOF statistic we have
adopted is the weighted, perpendicular color-distance residual.

The perpendicular color-distance is a hyper-dimensional distance in
color space between an instrumental data point and the nearest point
on the standard locus line.  Our standard locus line is a collection
of closely-spaced data points, so we make numerical approximations to
calculate the distance.  As we make explicit below, we first compute
all possible distances between the instrumental colors and the points
on the standard locus line, then we find the minimum distance for each
instrumental data point.  Finally we sum the result.

Consider instrumental color data points $\color_\alpha$, where
$\alpha$ indexes the instrumental color data vector of each star.  We
compute the vector distances between every $\color_\alpha$ and every
data point on our standard locus line, $\color_{0\beta}$, where
$\color_{0\beta}$ is also an array of vectors and $\beta$ varies along
the standard locus line.  These distances are
$\boldsymbol{d}_{\alpha\beta} \equiv \color_\alpha-\color_{0\beta}$,
where $\boldsymbol{d}_{\alpha\beta}$ is a tensor of all possible
distances between the instrumental colors and the points on the
standard locus line.  

We weight the distances either by the number $1$ or by the color
measurement uncertainty in the direction of the line connecting each
pair of points $\alpha$-$\beta$.  Whether to weight by unity or by
color uncertainties is at the discretion of the user.  We take the
norms of each weighted distance vector, so our weighted distance
measure is
\begin{equation}
  |\boldsymbol{d}_{\alpha\beta}^{\mathrm{w}}| = \frac{|\boldsymbol{d}_{\alpha\beta}|}{|\boldsymbol{\sigma}_{\alpha}\cdot \hat{\boldsymbol{d}}_{\alpha\beta}|}.
\end{equation}
Here the dot product $\cdot$ is taken between the vector of data
uncertainties $\boldsymbol{\sigma}_{\alpha}$ and the unit vector
that lies along the pairs of points
$\hat{\boldsymbol{d}}_{\alpha\beta}$.  The quantity
$\boldsymbol{\sigma}_{\alpha}\cdot
\hat{\boldsymbol{d}}_{\alpha\beta}$ is the uncertainty projected along the line that connects each pair of points.
% sorry but where did the pairs come from? 

Then, for each $\alpha$, we select the one $\beta=\beta^*$ such that
$|\boldsymbol{d}_{\alpha\beta^*}^{\mathrm{w}}|=\min(|\boldsymbol{d}_{\alpha\beta}^{\mathrm{w}}|)$.
This is the closest distance to the standard locus line for the
instrumental data point $\alpha$.  We do this for all data points
$\alpha$, and sum the results as
\begin{equation}
  T = \sum_\alpha |\boldsymbol{d}_{\alpha\beta^*}^{\mathrm{w}}|,
\end{equation}
noting that there is one $\beta^*$ for each $\alpha$.  This is an
estimate of the weighted distance residual between all instrumental data points
and the standard locus line---a scalar.  

Our \slr\ implementation produces best-fit calibration parameters by
varying $\zptcolor$ and adding it to the instrumental colors, then
recomputing $T$, and repeating this process until $T$ is minimized.
We use the amoeba downhill simplex method \citep{bib:amoeba}.

Our statistic was chosen for robustness.  $T$ would be equivalent to
the $\chi^2$ statistic if the distances were squared and weighted by
the inverse variance\footnote{In this case our \gof\ is closely
  related to the 7-dimensional color distance of \citet{bib:covey}.
  The difference is that we weight by the errors in the direction of
  the line connecting the data and the nearest point on the standard
  locus line.}, but the square puts undue weight on statistical
outliers.  Likewise, using uniform weights of unity, the residual of
square perpendicular distances becomes equivalent to the so-called
``total least-squares'' statistic.  This similarly gives undue weight
to {\it non}-statistical outliers.  We wish to apply \slr\ to any data
at the telescope, on the fly, so robustness to outliers is critical.

To insure doubly against outliers, we perform \slr\ twice.  The first
iteration gives rough estimates of best-fit calibration parameters,
allowing us to excise stars we deem to lie too far away from standard
locus line.  We typically cut stars with color distance
$|\boldsymbol{d}_{\alpha\beta^*}|>6|\boldsymbol{\sigma}_{\alpha}\cdot
\hat{\boldsymbol{d}}_{\alpha\beta^*}|$ away from our standard locus line
\citep[a figure informed by][]{bib:covey}, and as well as those with
$|\boldsymbol{d}_{\alpha\beta^*}|>1\,\mathrm{mag}$.  We then perform \slr\
once more on the cleaned data.

\subsection{Uncertainty Estimation}
\label{sec:errors}

We estimate all uncertainties on best-fit parameters with the
bootstrap method \citep{bib:efron}.  At each bootstrap iteration $i$
we resample $N_{\mathrm{stars}}$ data points $N_{\mathrm{stars}}$ times with replacement
to obtain a bootstrap sample, and perform the \gof\ optimization
described above on the new sample.  This gives best fit parameters
$\zptcolor_i$, which we record.  We repeat this
$N_{\mathrm{boot}}\sim100$ to $1000$ times, obtaining
$N_{\mathrm{boot}}$-size arrays of each vector $\zptcolor_i$.  The mean
of each bootstrap distribution is an estimate of the sample mean of
each parameter.  The standard deviation of the bootstrap distribution
is an estimate of the standard error on the mean.

This approach means that to estimate uncertainties we repeat the entire
stellar locus regression $N_{\mathrm{boot}}$ times.  This can be
time-prohibitive for large samples, but we argue it is not always
necessary to repeat the error estimation after it has been performed
once.  If different stellar images are taken from the same instrument
under similar circumstances, and if the bootstrap distributions are
roughly normal, then errors for different size data sets can be
estimated with rescaling by $\sqrt{N_{\mathrm{stars}}}$.  This estimate may break down if
those assumptions do not apply, but we argue $\sqrt{N_{\mathrm{stars}}}$ rescaling can
give errors to a factor of a few for $N_{\mathrm{stars}}$ that are not wildly
different.
We now describe the full algorithm we have implemented to produce
calibrated colors from flatfielded images.

\subsection{An Algorithm}
\label{sec:algorithm}

\begin{figure}
\plotone{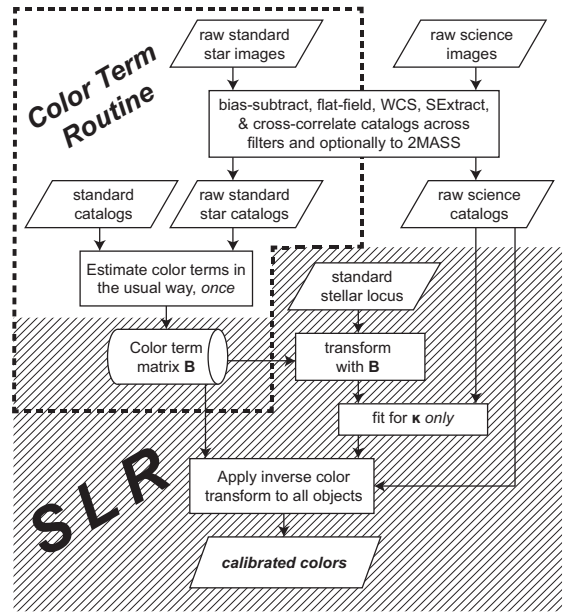}
\caption{ \slr\ flow chart for calibrating colors.  The hashed region
  denotes parts of the algorithm that are unique to \slr, while the
  non-shaded region shows steps that are more traditional.  The dotted
  region denotes the color term estimation routine, which need only be
  performed once per detector. }
\label{fig:algorithm}
\end{figure}

We have developed an optimized algorithm that produces \slr-calibrated
colors from flatfielded images, essentially in real time.  It is schematically
outlined in Figure \ref{fig:algorithm}.  The data undergo standard
preprocessing, color terms are determined in the normal way from
standard star frames {\it once}, and \slr\ fitting is done
subsequently on all science frames.

\slr\ requires single-epoch multiband observations of any number of
science fields and one standard star field.  All images are
bias-subtracted, flat-fielded, WCS-registered, and Source Extracted.
The resulting instrumental catalogs from each passband are cross-correlated
with one another to find the unique objects, and point-sources are
identified, for example, using Source Extractor's {\sc class\_star}
parameter.  All stars are required to be unambiguously identifiable in
all bands, and at all stages we only consider those with
signal-to-noise $>10$.

Instrumental color terms are then determined once only from the
standard star field, in the traditional way.  For example, following
the equations of Appendix \ref{app:corresp} we estimate the $g$-band
color term $b_g$ as the slope of the best-fit line to $g-g_0$ vs.\
$g_0-r_0$, and so forth for the rest of the passbands.  The color
terms are stored as the color term matrix $\colormatrix$.

With color terms in hand, real-time \slr\ calibration of any number of
science fields can be undertaken.  The color terms are first applied
to the standard stellar locus with the transform $(\identity+\colormatrix)$,
bringing the standard locus into {\it standardized instrumental color
  space}.  In this space, the instrumental stellar colors from the flat-fielded
science images are regressed to the standard locus, leaving only the
color translation vector $\zptcolor$ free.  This stage can be executed
as soon as flat-fielded multiband images are in hand.  Flat-fielding
and object cataloging can be performed at the telescope during
observation, so Stellar Locus Regression is viable as a real-time
calibration technique.

The instrumental science catalog of both point-like and extended
sources is calibrated by applying the inverse color transformation,
\begin{equation}
  \label{eqn:invtransform}
  \color_0 = (\identity + \colormatrix)^{-1}\left(\color-\zptcolor\right).
\end{equation}
$\colormatrix$ was determined from the standard star observations, and
$\zptcolor$ was obtained from the \slr.  The transform $(\identity +
\colormatrix)$ is invertible under normal circumstances of small color
terms (see Appendix \ref{app:corresp}).

As outlined in \S\ref{sec:2mass}, \slr\ generates calibrated
magnitudes using \tmass\ entries using exactly the color calibration
process described above, with judicious choices of color vector
entries.  The additional step of matching one's instrumental objects to
\tmass\ lasts seconds to minutes for typical catalogs containing of
order tens to hundreds of stars.

We apply this algorithm to real data in a series of test, presented in
the following section.

%%%%%%%%%%%%%%%%%%%%%%%%%%%%%%%%%%%%%%%%%%%%%%%%%%%%%%%%%%%%%%%%%%%%%%%%%%%%%%%%

\section{First Tests}
\label{sec:tests}

We employed the algorithm of \S\ref{sec:algorithm} on existing,
calibrated \sdss\ data and on new $griz$ data we have acquired on the
$6.5\,\mathrm{m}$ Magellan telescopes using the \imacs\
\citep{bib:imacs,bib:magellan2008} and \ldss\
\citep[see][]{bib:magellan2008} instruments in imaging mode.  We designed a series
of tests to isolate the effect of Galactic extinction and airmass and
to generally assess the reproducibility of colors using \slr.
Finally, we used \slr\ in a measurement of redshifts of $11$ galaxy
clusters using colors alone.

\subsection{Galactic Extinction}
\label{sec:galext}

The majority of \sdss\ stars at high Galactic latitude are expected to
be behind the dust (\S\ref{sec:stellarlocus}), so we designed a simple
test of \slr\ to measure the extinction directly.  We applied \slr\ to
ubercalibrated \citep{bib:ubercal} \sdss\ stars subject to varying
degrees of predicted reddening, and compared the results to SFD
expectation.  Ubercal \sdss\ data are calibrated to the top of the
atmosphere, in front of the Galactic dust.  We therefore expect the
best-fit color vector $\kappa$ to be equal to the reddening vector,
especially at high Galactic latitudes, and we expect the canonical
$R_V=3.1$ extinction law to hold.

We began by querying photometric quality ubercalibrated \sdss\ point
sources from the CasJobs web
server\footnote{\url{http://casjobs.sdss.org/CasJobs/}}.  Photometric
quality ubercalibrated colors are accurate over a wide area of the
sky, with $1.5\%$ estimated uncertainty, so this is the ideal \sdss\
data set to perform our test on, in terms of size and quality.  We
queried stars in small ranges of SFD-predicted extinction and
performed multiple such queries over a wide range of mean extinctions.
The results were samples of hundreds of stars each in finite Galactic
extinction bins over a wide range of SFD dust columns.  We selected
stars with signal-to-noise ratio $>10$ in all bands to minimize
uncertainty from poor color measurement.

\begin{figure}
\plotone{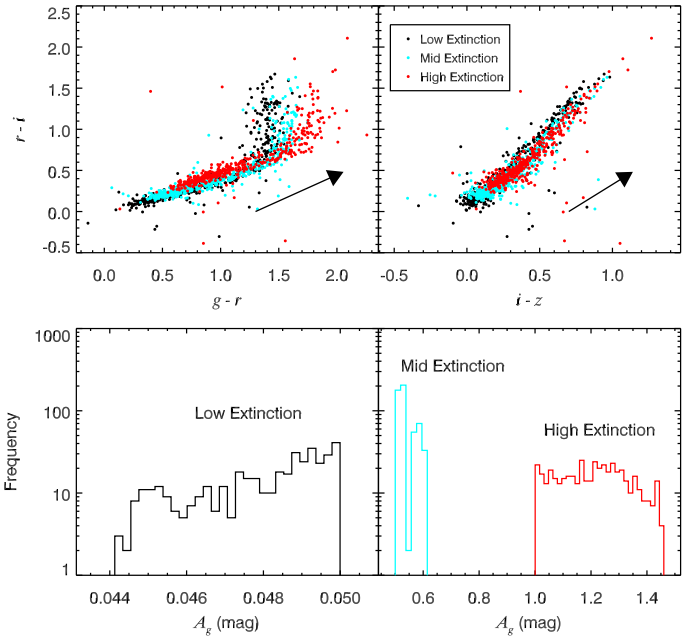}
\caption{ Three samples of \sdss\ stars with different ranges of
  predicted Galactic extinction (low: $A_g\sim0.05\,\mathrm{mag}$,
  intermediate: $A_g\sim0.5\,\mathrm{mag}$, and high:
  $A_g\sim1.2\,\mathrm{mag}$).  {\it Top panels}: color-color stellar loci,
  shown with the canonical reddening unit vector in three-dimensional
  color space.  {\it Bottom panels}: histograms for predicted $g$-band
  Galactic extinction. }
\label{fig:galext}
\end{figure}

Figure \ref{fig:galext} displays three of these data sets.  Low
($A_g\sim0.05\,\mathrm{mag}$), intermediate ($A_g\sim0.5\,\mathrm{mag}$),
and high ($A_g\sim1.2\,\mathrm{mag}$) Galactic extinction ranges were
queried, as shown in the figure.  Histograms of the SFD-predicted
extinction in the $g$-band for each star show that these extinction
distributions are localized and well separated.  The gross effect of
Galactic reddening is readily apparent in color-color diagrams (top
panels).

In total we acquired $7$ such data sets, sampled from the northern and
southern Galactic hemispheres, probing mean $A_g$ between $0.05$ and
$1.2\,\mathrm{mag}$.  By nature of the \sdss\ database organization and
the available data itself, the data sets happen to come from fields
that are localized to $\sim1\deg^2$, except for one. The
Galactic coordinates are given in Table \ref{tab:galext}.  The last
field in Table \ref{tab:galext} shows a wide range of stellar
coordinates because the CasJobs query, by chance, returned stars
spread over a few disjointed regions of the sky.

\begin{deluxetable}{crr}
%\tabletypesize{\small} % 11pt
\tabletypesize{\footnotesize} % 10pt
%\tabletypesize{\scriptsize} % 8pt
%\rotate
  \tablewidth{0pt} 

  \tablecaption{Galactic Extinction Fields.\label{tab:galext} }

  \tablehead{ 
    \colhead{Mean $A_g$ (mag)}&%\tablenotemark{a}} & 
    \colhead{$\ell$ (deg)\tablenotemark{a}} &
    \colhead{$b$ (deg)\tablenotemark{b}}
}

\startdata

0.048 & $57.9\pm1.0$ & $37.9\pm0.5$ \\
0.538 & $15.8\pm0.2$ & $29.0\pm0.2$ \\
1.210 & $11.4\pm0.1$ & $36.2\pm0.1$ \\
0.090 & $130.6\pm0.9$ & $-49.0\pm0.2$ \\
0.268 & $115.2\pm0.4$ & $-48.7\pm0.1$ \\
0.727 & $149.8\pm0.4$ & $-46.6\pm0.1$ \\
1.068 & $56.6\pm33.9$ & $-37.5\pm3.1$ \\

\enddata

\tablenotetext{a}{Mean and standard deviation of Galactic longitudes
  for stars in the sample.}
\tablenotetext{b}{Mean and standard deviation of Galactic latitudes
  for stars in sample.}

\end{deluxetable}

\begin{figure}
\plotone{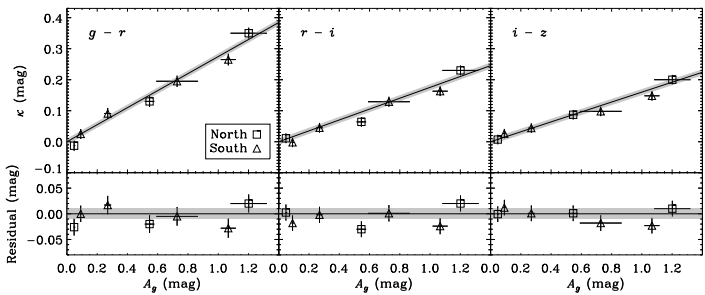}
\caption{ Best-fit color shifts $\zptcolor$ vs.\ predicted Galactic
  extinction in the $g$-band, and residuals.  The solid lines are the
  SFD-predicted reddening corresponding to the $g$-band extinction
  given on the abscissa.  Northern and southern Galactic hemisphere
  fields are indicated, and the $\pm10\,\mathrm{mmag}$ region is shaded.
}
\label{fig:lowmidhigh}
\end{figure}

We performed Stellar Locus Regressions on each data set.  We fixed the
color term matrix $\colormatrix$ to zero because this test involved
comparing a standard stellar locus standardized to the \sdss\
photometric system, to colors generated by the \sdss\ instrument
itself.  After our restrictions on the data were enforced (see
\S\ref{sec:gof}), between $380$ to $490$ stars from each data set were
ultimately used in the fits.  The best-fit color shifts $\zptcolor$
that resulted are directly compared to the SFD-predicted reddening in
Figure \ref{fig:lowmidhigh}.  The expectation, if our hypotheses hold,
is that $\zptcolor$ will be equal to the reddening vector.

The solid lines in Figure \ref{fig:lowmidhigh} show the {\it
  predicted} color-reddening values $A_n-A_m$, $n,m\in\{g,r,i,z\}$
corresponding to the predicted $g$-band extinction given on the
abscissa, assuming the canonical $R_V=3.1$ reddening law.  The error
bars for the abscissa are the standard deviation of predicted
extinction values for the corresponding data set.  Error bars for the
ordinate are the standard error on the mean predicted reddening added
in quadrature to the bootstrap errors of \slr\ and the $1.5\%$ color
uncertainty limit to which ubercalibrated colors are subject.

The best-fit color shifts from \slr\ are consistent with Galactic
reddening within the errors for these data, with residual RMS of
$19\,\mathrm{mmag}$ in the color $g-r$, $18\,\mathrm{mmag}$ in $r-i$, and
$13\,\mathrm{mmag}$ in $i-z$.  The maximum residual disagreement is
$30\,\mathrm{mmag}$ in $g-r$ and $r-i$, and $19\,\mathrm{mmag}$ in $i-z$.

\citet{bib:ivezic_stripe82} performed a similar analysis in \sdss\
Stripe 82 with their ``stellar locus method,'' showing maximum
sensitivity to the magnitude of SFD extinction at roughly the
$20\,\mathrm{mmag}$ level at high Galactic latitudes.  Their
measurements showed disagreement with SFD prediction at Galactic
latitudes $|b|\sim40$--$50\deg$ and below (their \S2.7.1).  Our
results probed latitudes in this range and lower, but in fields from
different Stripes, so our results are not inconsistent with theirs.
In other tests (\S\ref{sec:colormag}), we observed Stripe 82 fields
where their stellar locus method failed, and we reproduced their
results.  Taking this information together, we conclude that stellar
locus methods {\it can} reproduce the $R_V=3.1$ reddening law in both
magnitude and direction, even through significant dust columns, so
dust thickness alone is not a good indicator of \slr\ reliability.

\subsection{Color and Magnitude Reproducibility}
\label{sec:colormag}

We tested the ability of \slr\ to reproduce colors and magnitudes
using data acquired at different telescopes.  We chose Stripe 82
fields because of the high quality, $1\%$ photometry available from
\citet{bib:ivezic_stripe82}.  We probed fields that contained among
the highest densities of stars in the entire Stripe 82 in order to
minimize statistical uncertainty.  The stars in these regions,
however, are known not to reproduce the SFD extinction prediction with
stellar locus methods \citep[\S2.7.1 of][]{bib:ivezic_stripe82}.  The
possible reasons cited were an invalid extinction law $R_V$, a drift
in mean metallicities, or dust that was both in front of and behind
the stars.  Nonetheless, we designed our experiment to control for
the anomalous dust, as described below, so that we could study in
isolation \slr's fundamental ability to reproduce colors and
magnitudes when applied to data from different instruments.

In 2008 we undertook observing programs at the $6.5\,\mathrm{m}$
Magellan telescopes using the \imacs\ instrument in imaging mode.  We
observed three, high stellar density Stripe 82 fields, which we label
S1, S2, and S3.  Our field S1 is centered at $(\ell,b)=(48.65151,
-26.04729)\deg$, S2 at $(\ell,b)=(51.86849, -33.74441)\deg$, and S3 at
$(\ell,b)=(50.43107, -30.24483)\deg$.  These fields are all at lower
Galactic latitude $|b|$ than those of \S\ref{sec:galext}, except for
the second field shown in Table \ref{tab:galext}.  Each \imacs\ CCD is
$4'\times8'$, arranged in two rows of four.  Just as we have described
in \S\ref{sec:algorithm}, we flat-fielded the data and produced
instrumental, multiband catalogs.  We associated stars between
passbands by requiring their positions to agree to better than $1''$
in radius.  We then found the corresponding photometry in the catalog
of \citet{bib:ivezic_stripe82} in the same way, selecting only those
stars with more than 4 SDSS observations per passband and with
signal-to-noise $>10$.

In order to compare the colors of stars directly, we treated the
\sdss\ catalog data as pseudo-\imacs\ observations: we extracted only
the stars that also appeared in the 8 \imacs\ CCDs, in all $griz$
bands.  This was repeated for the three fields.  Because we wished to
compare \slr\ magnitudes in addition to colors, we also found the
\tmass\ $J$-band photometry for the stars.  Of order $10$--$100$ stars
per CCD per field were matched across the \imacs, \sdss, and \tmass\ data sets.

Following the procedure outlined in \S\ref{sec:colormatrix} and
\S\ref{sec:algorithm}, we measured the \imacs\ color terms from the
\sdss\ standard stars.  We exploited the fact that we observed $3$
standard fields, and measured color terms independently in all of
them, one CCD at a time, taking the average of the result.  We applied
the same mean color term correction to all CCDs.  This color term
procedure gave smaller color and magnitude residuals than (1) using
different mean color terms for each CCD and (2) using separate color
terms for each field {\it and} CCD.

First, we ran \slr\ to calibrate only the colors.  We controlled for
the anomalous stellar locus shifts by regressing both the calibrated
\sdss\ data and our instrumental colors to the standard locus line.
The best-fit $\zptcolor$ for the \sdss\ data showed color translations
that disagreed with SFD prediction by $\sim50$--$100\%$, consistent
with the findings of \citet{bib:ivezic_stripe82}.  By applying \slr\
to both the instrumental and \sdss\ data sets, we effectively canceled
the anomaly by subtraction.

In a separate, subsequent step we calibrated the photometry using
\slr, fixing the optical-only color shifts to those measured in the
first step, as described in \S\ref{sec:2mass} and
\S\ref{sec:algorithm}.  We solved for the \slr-standardized photometry
as
\begin{subequations} 
\begin{align}
  g_0 & = g-\kappa_{gJ}-b_g(g_0-r_0) \\
  r_0 & = r-\kappa_{rJ}-b_r(r_0-i_0) \\
  i_0 & = i-\kappa_{iJ}-b_i(i_0-z_0) \\
  z_0 & = z-\kappa_{zJ}-b_z(i_0-z_0)
\end{align} 
\end{subequations}
where the color terms $b_n$ were the measured values for the \imacs\
catalogs, and zero for the \sdss\ catalogs.  For the colors on the
right-hand side, we used the \slr\ calibrations from the first,
optical-only color iteration.  The $\kappa_{gJ}$, $\kappa_{rJ}$, {\it
  etc}, were those obtained from the second, optical-infrared
magnitude iteration of \slr.

% Resulting best-fit $\zptcolor$ are shown in Figure \ref{fig:galext2}.
% \begin{figure}
% \plotone{fig/color_res_vs_galext.eps}
% \caption{ \footnotesize Best-fit $\zptcolor$ from SDSS Stripe 82 data
%   for each mock-CCD vs.\ SFD Galactic extinction.  The solid lines are
%   as in Figure \ref{fig:lowmidhigh}.  The discrepancies here are
%   consistent those shown by \citet{bib:ivezic_stripe82} for the same
%   fields. }
% \label{fig:galext2}
% \end{figure}

\begin{figure}
\plotone{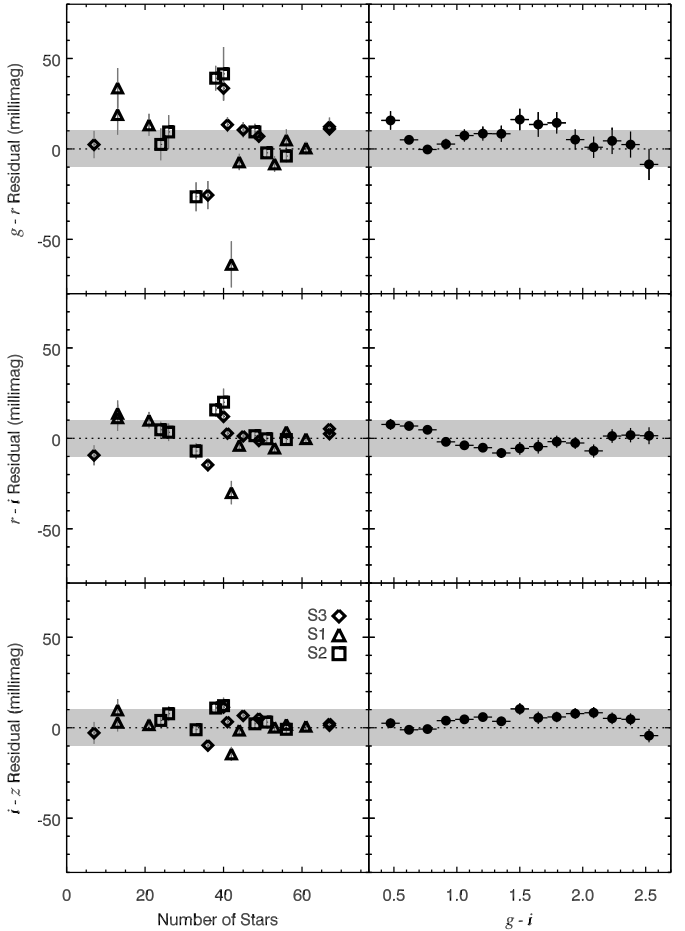}
\caption{ \slr\ color reproducibility, where \slr\ was applied both to
  \sdss\ and new \imacs\ observations.  Color residuals are plotted
  vs.\ number of stars used per \slr\ ({\it left panels}) and vs.\
  true $g-i$ color ({\it right panels}).  Each data point on the left
  represents an \slr\ calibration done on one CCD, and residuals vs.\
  $g-i$ are averaged over the union of all stellar residuals across
  all fields and CCDs.  The $\pm10\,\mathrm{mmag}$ region is shaded. }
\label{fig:res_s82_c}
\end{figure}

\begin{figure}
\plotone{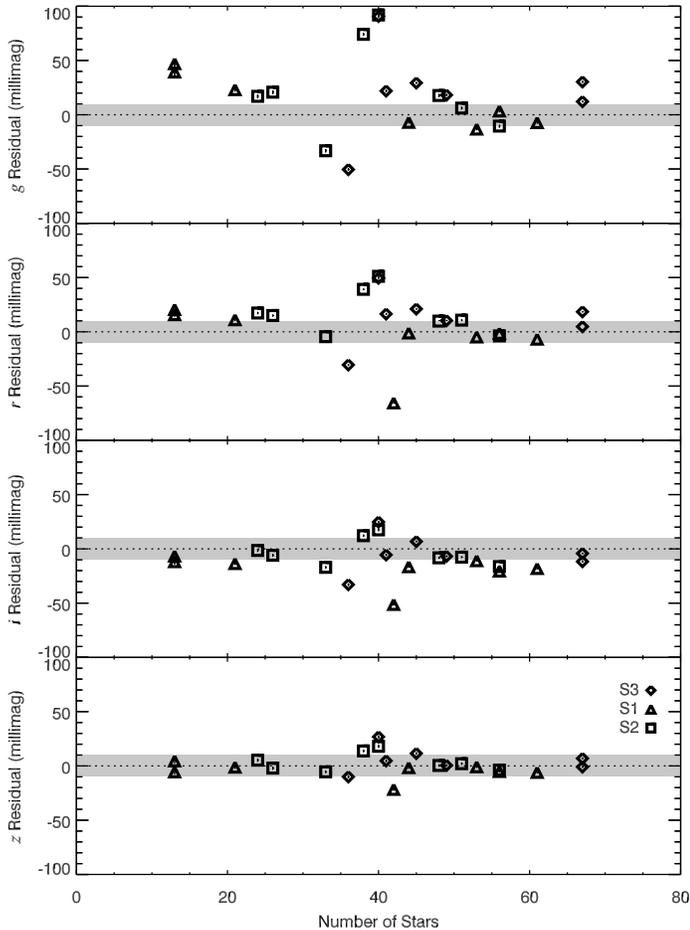}
\caption{ \slr\ magnitude reproducibility, where \slr\ was applied
  both to \sdss\ and new \imacs\ observations.  A \tmass\ \slr\
  magnitude calibration was performed, again on both data sets as in
  Figure \ref{fig:res_s82_c}, with one data point per CCD.  The
  $\pm10\,\mathrm{mmag}$ region is shaded. }
\label{fig:res_s82_m}
\end{figure}

Figures \ref{fig:res_s82_c} and \ref{fig:res_s82_m} show the resulting
color and magnitude residuals.  The figures show a lack of pronounced
systematic error in \slr\ colors as a function of effective
temperature, $g-i$.  We assess that colors calibrated purely with
\slr\ are reproducible between the \sdss\ and the \imacs\ instruments
to $(18,6,5)\,\mathrm{mmag}$ in the colors $(g-r,r-i,i-z)$.
Magnitudes obtained only with \slr, using \tmass, are reproducible
between \imacs\ and \sdss\ to $(44,25,19,9)\,\mathrm{mmag}$ in the
$(g,r,i,z)$ passbands.  These numbers do not include the Galactic dust
uncertainty because we applied \slr\ to both data sets, but they do
necessarily include the intrinsic airmass correction that \slr\ makes.
Our following test isolated completely the effect of the atmosphere
using \imacs\ data alone.

\subsection{The Atmosphere}
\label{sec:atmext}

We again tested the reproducibility of \slr-calibrated colors, this
time isolating and varying the effect of atmospheric extinction.  An
advantage to the \slr\ approach is that is does not assume any
particular functional dependence of color variation with airmass
(\S\ref{sec:kappa}).  We therefore expect precise airmass corrections
from \slr.

We isolated the effect of the atmosphere by performing \slr\ on the
{\it same sets of stars} observed through $2$ different airmasses
using the same instrument, and compared directly the \slr-calibrated
colors of the matched stars.  We observed the \imacs\ fields described
in \S\ref{sec:colormag} through multiple airmasses with the $griz$
passbands.  We included two additional fields, S5 at
$(\ell,b)=(122.46159, -63.21097)\deg$ and CL1 at $(\ell,b)=(96.07818,
-61.31761)\deg$---the latter of which we observed at three different
airmasses.  The field S5 is in Stripe 82, but did not contain enough
matches between \imacs, \sdss, and \tmass\ to include in the analysis
of \S\ref{sec:colormag}.  The field CL1 is not in the \sdss\
footprint.

The data preprocessing was identical to that of \S\ref{sec:colormag},
except here we applied the nonzero color term corrections to all
\imacs\ data, and we did not cross-correlate with \tmass\ as we only
wish to examine colors.

\begin{figure}
\plotone{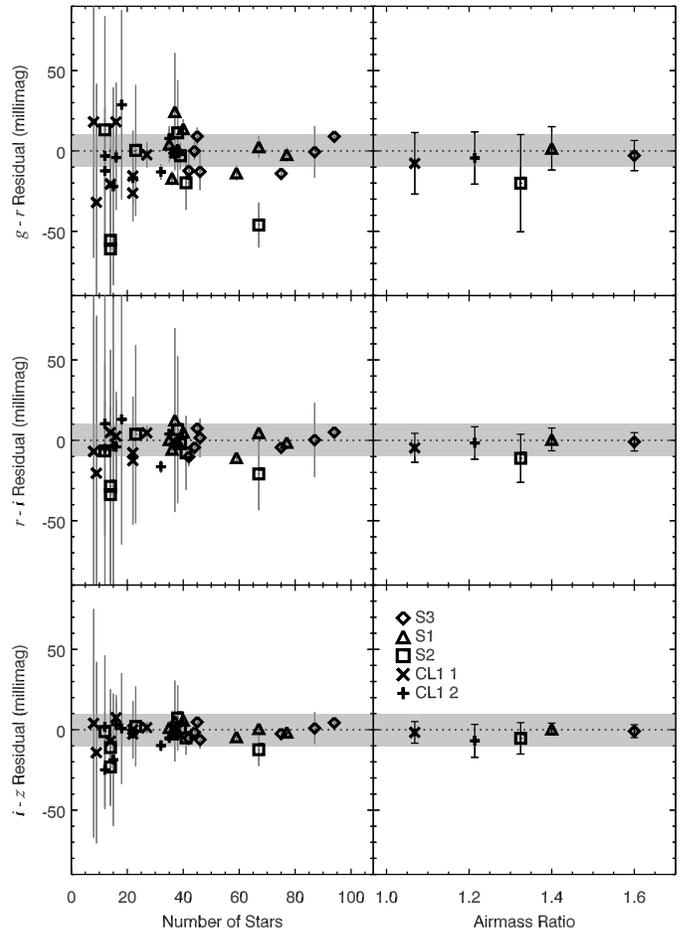}
\caption{ Color residuals between \slr\ calibrations of fields imaged
  with IMACS at different airmasses.  The $\pm10\,\mathrm{mmag}$ region
  is shaded.  {\it Left hand plots:} Residuals vs.\ number of stars
  used to perform each \slr.  {\it Right hand plots:} Mean of
  residuals over the $8$ \imacs\ CCDs vs.\ airmass ratio between the
  two observations, and error bars are the RMS of residuals per field.
}
\label{fig:res_am}
\end{figure}

Figure \ref{fig:res_am} shows the results.  The left panel shows the
weighted mean residuals of stellar colors per CCD per field, with
bootstrap errors, plotted against the number of useful stars extracted
from each CCD.  The right hand panel shows the mean of each set of $8$
data points per field, with error bars reflecting the RMS scatter,
plotted vs.\ the airmass ratio of the two observations.  CL1 was
observed at three airmasses, and the highest two airmasses are each paired
with the lowest airmass in this figure.

This shows that \slr\ colors are reproducible over a wide range of
airmasses, with residual RMS scatter in $(g-r,r-i,i-z)$ of
$(14,8,5)\,\mathrm{mmag}$ when the number of stars used is $>30$, and
$(20,10,8)\,\mathrm{mmag}$ overall.  Color errors do not correlate well with
increased airmass.  As is expected, there is a dependence of the
color errors on the number of stars used per regression.  The reduced
$\chi^2$ of the fits are roughly in the range $[1,3]$, with larger
values for bluer bands.  The larger $N_{\mathrm{stars}}$ data exhibit
slightly larger $\chi^2$, and smaller $N_{\mathrm{stars}}$ data exhibit
slightly smaller $\chi^2$.  This suggests that the bootstrap slightly
overestimates errors at small $N_{\mathrm{stars}}$, and slightly
underestimates errors at larger $N_{\mathrm{stars}}$.  This may be due
to mildly non-Gaussian distributions.  Overall, these results suggest
that \slr\ systematic errors are not dominated by airmass issues for
the fields we tested, but instead are dominated by the statistical
noise inherent to the method as applied to the particular data we have
used it on here.

As our final test, we deployed \slr\ colors in a photometric redshift
measurement, as distance estimation is one of the prime uses of
accurately calibrated colors.

\subsection{Galaxy Cluster Redshifts}
\label{sec:photoz}

We used \slr-calibrated colors to recover the redshifts of galaxy
clusters.  As part of the 2008 observing program we mentioned in
\S\ref{sec:colormag} and \S\ref{sec:atmext}, we observed known galaxy
clusters from the REFLEX catalog \citep{bib:reflex}, primarily a
subset that are also Abell clusters \citep{bib:abell}.  These
observations were acquired with the \ldss\ camera at the Magellan
$6.5\,\mathrm{m}$ telescopes \citep[see][]{bib:magellan2008} in imaging
mode.  Because the redshift of the clusters are known, this serves as
a test of our color calibrations within the context of a full
scientific analysis.

We estimated \ldss\ color terms using standard star fields.
Individual \slr\ calibrations were then obtained directly from each of
the single-epoch $griz$ cluster fields, and the $\colormatrix$ and
best-fit $\zptcolor$ applied to the galaxy colors without applying any
further corrections for airmass or Galactic extinction.  We then
selected the brightest red-sequence cluster galaxies from
color-magnitude diagrams, spanning $\sim 1$--$2\,\mathrm{mag}$ fainter
than the brightest cluster galaxy (BCG) magnitude.  This typically
included of order $10$ galaxies.  We calculated the weighted mean of
their $g-i=(g-r)+(r-i)$ colors and estimated the cluster redshift with
the empirical color-to-redshift tables of \citet{bib:lopes}.  For the
low redshift range we probed, the $g-i$ color of red cluster galaxies
varies rapidly and monotonically with redshift and is an ideal
redshift estimator.

Table \ref{tab:clusters} summarizes the galaxy clusters we targeted,
along with their Galactic coordinates and our results.  The Abell
systems 3693 and 3738 have been identified by previous authors as
multi-redshifts systems.  We were able to isolate the sub-systems
using color-magnitude diagrams and image inspection, and we estimated
their redshifts independently.

\begin{deluxetable*}{lllllllll}
%\tabletypesize{\small} % 11pt
%\tabletypesize{\footnotesize} % 10pt
%\tabletypesize{\scriptsize} % 8pt
%\rotate
\tablewidth{0pt}
\tablecaption{Summary of the Cluster Data and Results.\label{tab:clusters} }

\tablehead{
   \colhead{ Name} &
   \colhead{References\tablenotemark{a}} &
   \multicolumn{2}{c}{Redshift} &
   \colhead{Airmass\tablenotemark{d}} &
   \colhead{Seeing\tablenotemark{e}} & 
   \colhead{$E_{g-i}$\tablenotemark{f} } &
   \colhead{$N$ stars\tablenotemark{g}} &  
     \colhead{$b$\tablenotemark{h}}  \\
   \colhead{~} & \colhead{~} & \colhead{Spec\tablenotemark{b}} & \colhead{\slr\tablenotemark{c}} & \colhead{~} &  \colhead{(arcsec)}  & \colhead{(mag)} & \colhead{~} & \colhead{(deg)} \\
}

\startdata

ABELL 3668 & $1,2$ & 0.1496 & $0.156 \pm 0.011$ & 1.9 & 2.0--3.0 & 0.083 & 58 &  $-32$ \\
ABELL 3675 & $1,2$ & 0.1383 & $0.150 \pm 0.012$ & 1.9 & 1.8--1.9 & 0.079 & 67 &  $-35$ \\
RXC J2023.4-5535 & $2$ & 0.232 & $0.238 \pm 0.012$ & 1.7 & 1.7--2.3 & 0.102 & 70 &  $-35$ \\
ABELL 3693 & $1,3,4$ & $0.091$ & $0.093 \pm 0.016$ & 1.5 & 1.8--3.0 & 0.060 & 62 &  $-35$ \\
ABELL 3693 & $1,4,5$ & $0.123 $ & $0.120 \pm 0.016$ & 1.5 & 1.8--3.0 & 0.060 & 62  & $-35$ \\
ABELL 3739 & $1,2$ & $0.1651$ & $0.177 \pm 0.010$& 1.3 & 1.6--2.1 & 0.054 & 63 &  $-42$ \\
ABELL 3739 & $1,6,7$ & $0.1786$ & $0.183 \pm 0.010$& 1.3 & 1.6--2.1 & 0.054 & 63 &  $-42$ \\
ABELL 3740 & $1,3$ & 0.1521 & $0.141 \pm 0.007$& 1.3 & 1.4--2.7 & 0.086 & 39 & $-42$ \\
ABELL 3836 & $1,3$ & 0.11 & $0.120 \pm 0.007$& 1.3 & 1.4--2.7 & 0.032 & 32 &  $-51$ \\
RXC J2218.6-3853 & $8$ & 0.1379 & $0.138 \pm 0.022$ & 1.5 & 1.2--1.9 & 0.026 & 12 &   $-56$ \\
ABELL 3866 & $1,9$ & 0.1544 & $0.156 \pm 0.008$ & 1.4 & 1.1--1.9 & 0.020 & 22 &  $-57$ \\
ABELL 3888 & $1,10$ & 0.152912 & $0.159 \pm 0.016$ & 1.4 & 1.3--1.9 & 0.029 & 28 & $-59$ \\
AM 2250-633 & $9,11$ & 0.2112 & $0.204 \pm 0.014$ & 1.2 & 1.1--1.4 & 0.046 & 26 &  $-49$ \\

\enddata

\tablenotetext{a}{References---(1) \citet{bib:abell}, (2) \citet{bib:reflex}, (3) \citet{bib:struble}, (4) \citet{bib:katgert}, (5) \citet{bib:zaritsky}, (6) \citet{bib:rbs}, (7) \citet{bib:schwope}, (8) \citet{bib:rass1}, (9) \citet{bib:degrandi}, (10) \citet{bib:pimbblet}, (11) \citet{bib:arp}.}
\tablenotetext{b}{Spectroscopic redshift reported by references.}

\tablenotetext{c}{Photometric redshift based on \slr\ color
  corrections, with standard errors on the mean redshift estimates of
  the brightest cluster member galaxies. }

\tablenotetext{d}{Airmass at which the images were obtained.}
\tablenotetext{e}{Range of seeing in the multiband images.}
\tablenotetext{f}{SFD-predicted Galactic reddening in $(g-i)$.}
\tablenotetext{g}{Number of stars used to determine the \slr\ color correction.}
\tablenotetext{h}{Galactic latitude of the cluster.}

\end{deluxetable*}

\begin{figure}
%\plottwo{clusterz_io.eps}{clusterz_err_vs_galext.eps}
\plotone{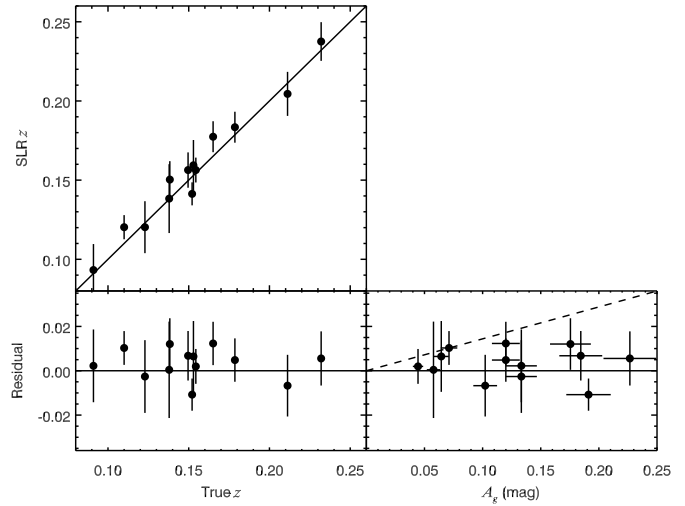}
\caption{ Photometric galaxy cluster redshifts from \slr-corrected
  colors vs.\ spectroscopic redshifts, and residuals.  {\it Right
    panel:} Redshift residuals vs.\ expected $g$-band Galactic
  extinction.  The red dashed line is the redshift error expected
  using colors that are not corrected for Galactic extinction.  There
  is no evidence for a redshift residual correlated with Galactic
  extinction, indicating that \slr\ makes an appropriate correction. }
\label{fig:clusterzio}
\end{figure}

The results for the $11$ clusters are plotted in Figure
\ref{fig:clusterzio}.  The $g-i$ cluster color errors SE$_{gi}$ were
estimated as the standard error on the mean $g-i$ color of the red
sequence galaxies we selected, and cluster redshift errors were taken
to be SE$_{gi}/3$, taken from the slope of the color-redshift relation
of \citet{bib:lopes}.  The reduced $\chi^2$ of residuals is $0.6$, so
this error slightly overestimates the scatter.  The residual cluster
redshift RMS is $\sigma_z=0.007$, or $\sigma_z/(1+z)=0.6\%$ for the
range of redshifts we measured.  We emphasize that this is the
estimated redshift error not per red galaxy, but per cluster, each of
which made use of an ensemble of red galaxies.

The $\sigma_z=0.007$ RMS error corresponds to $g-i$ color errors of
$\sim20\,\mathrm{mmag}$, again estimating from the roughly linear
relationship between redshift and $g-i$.  This level of residual error
is roughly in accord with the \slr\ systematic errors due to Galactic
dust and the atmosphere, estimated in \S \ref{sec:galext}, \S
\ref{sec:colormag}, and \S \ref{sec:atmext}, and the residuals are
inconsistent with erroneous Galactic dereddening.  If Galactic
extinction were completely unaccounted for by our calibration method,
redshifts would lie along the dashed line of the right-hand panel of
Figure \ref{fig:clusterzio}, assuming no other systematic effects.
Because our redshift residual trend is not consistent with the dashed
line, these results provide further evidence (see \S\ref{sec:galext})
that \slr\ accounts for Galactic extinction in these fields.

% \subsubsection{What about luminosity priors for photo-zs?}

% While many of the more popular photo-z codes use exclusively color
% information for the galaxies (\textit{kphotoz} \citep{bib:kcorrect};
% \textit{Hyper-z} \citep{bib:hyperz}; \textit{ANNz} \citep{bib:annz}),
% galaxy luminosity and/or surface brightness can be helpful in breaking
% ``catastrophic'' failures of the photo-z technique (refs).  Galaxies
% at a fixed redshift span a wide range of intrinsic luminosities
% (Press-Schec).  This implies that requirement for zeropoint
% calibration for luminosity prior is {\bf WAY} less demanding than what
% is needed to get precise colors by subtracting two photometrically
% calibrated magnitudes.

% So this should be easy, like 0.1 mag is probably OK. Can use crude
% calibrator in virtually any band to do this well. Even perhaps
% bootstrap off of \tmass.

%%%%%%%%%%%%%%%%%%%%%%%%%%%%%%%%%%%%%%%%%%%%%%%%%%%%%%%%%%%%%%%%%%%%%%%%%%%%%%%%

\section{Discussion}
\label{sec:discussion}

We have provided a new way to calibrate instrumental photometry.  It differs
substantially from traditional techniques, and so a side-by-side
comparison of \slr\ with traditional approaches is in order.  This the
first item of discussion.  We then list in detail the unique
advantages and limitations of \slr, and complete our discussion with
tasks for the future.

\subsection{\slr\ vs.\ Traditional Photometric Calibration}

\subsubsection{Traditional Photometric Calibration}

The traditional path to calibrated colors has been to first calibrate
magnitudes, then to subtract them.  Magnitudes are calibrated by tying
instrumental flux to photometric standard stars, which most often
requires additional exposure time investment outside one's field of
scientific interest.  The resulting zeropoints must be interpolated in
space and time to the science fields, but only after extinction by the
atmosphere is modeled and measured---which necessitates further
standard field observations over a wide range of zenith angles.

The analysis of these data to extract calibrated colors traditionally
involves the steps of:
\begin{enumerate}

\item{} Bias subtraction and flat-fielding;

\item{} Extracting instrumental magnitudes for all images, in all
  bands;

\item{} Using either observations of standard stars or program objects
  over a substantial span of zenith angles, in each band, to determine
  atmospheric extinction coefficients, and correcting each observation
  for atmospheric extinction;

\item{} Assessing whether the atmospheric conditions during
  observation were photometric, and thus whether the zeropoint and
  atmospheric extinction interpolations are valid;

\item{} Correcting for the frame-by-frame difference between
  PSF-fitting magnitudes and aperture photometry;

\item{} Determining and correcting for any differential sensitivity
  mismatch between the instrument and the desired photometric system,
  and applying these color term corrections to the photometry;

\item{} Determining Galactic extinction in the direction of the
  observation from some external source such as SFD, and making an
  appropriate correction to all magnitudes; and

\item{} Generating colors of sources of interest by subtracting these
  fully calibrated apparent magnitudes.

\end{enumerate}

The result of this procedure is a source catalog that contains
magnitudes and derived colors.  Besides being costly in telescope
time, this procedure is also suffers from extra uncertainty from the
explicit mathematical modeling of atmospheric attenuation.  Moreover,
the individual magnitudes must each be calibrated to a better
fractional precision than the desired color error.  In the future,
when high accuracy all-sky photometric catalogs of faint sources are
readily available in all bands of interest, this process will be
somewhat simplified.  In the meantime we offer the \slr\ approach.
\slr\ renders the preliminary magnitude calibration step unnecessary,
instead allowing for the direct determination of colors of all objects
of interest.  \slr\ also provides a way to calibrate the photometry by
tying the magnitude scale to the all-sky \tmass\ catalog, again
without the need for extra standard star observations.

\subsubsection{\slr\ Photometric Calibration}

The \slr\ approach matches the distribution of stars in the
instrumental color-color space to a standard stellar locus, allowing
us to replace the traditional analysis path with a streamlined set of
steps.
\begin{enumerate}

\item{} For each previously uncharacterized instrument, we observe
  only one field that contains stars whose calibrated magnitudes are
  known.  This allows us to establish, in the traditional way, the
  instrumental color terms that arise from filter and detector
  differences.

\item{} All subsequent images are bias subtracted and flat-fielded.

\item{} Instrumental magnitudes are extracted for objects in each
  image.

\item{} Objects are immediately cast in instrumental color-color
  space, and we determine the transformations needed to bring the
  instrumental stellar locus into agreement with the universal,
  calibrated color properties of stars.

\item{} This color transformation is then applied to all photometry,
  producing calibrated colors for all measured objects.

\item{} If desired, calibrated photometric zeropoints are determined
  by bootstrapping the photometry to any stellar photometric catalog
  that overlaps the program fields.  At the time of this writing,
  \tmass\ is the obvious choice.

\end{enumerate}
The result is a catalog of colors, and optionally magnitudes that are
calibrated in a separate step and whose uncertainties do not factor
into those of the colors.  As a fundamentally different calibration
technique, \slr\ has unique (in)sensitivities to astrophysical and
instrumental effects.  We break out the unique advantages and
limitations these afford.

\subsection{Advantages}

\paragraph{\slr\ corrects for atmospheric extinction, even if
  time-variable.}

Atmospheric extinction distorts the apparent colors of celestial
sources, compared to what would be observed at the top of the
atmosphere. To first approximation, this produces a translation of
stars in color-color space. The atmospheric transmission function is
expected to change as a function of time, zenith angle, and azimuthal
angle as parcels of aerosols and water vapor travel and evolve in the
sky \citep[cf][]{bib:atmos}.  Traditional airmass and zeropoint
interpolations cannot account for this, unless specially designed
systems to monitor atmospheric conditions along the line of sight are
deployed in parallel during observation.  Moreover, the $\sim X^{1/2}$
airmass dependence of saturated water lines in the $z$-band is simply
not modeled by most observers (\S\ref{sec:kappa}).  \slr\ is unique
because it naturally corrects for all of these additive effects at
once with $\zptcolor$, and is therefore insensitive to any additive
mis-modeling of atmospheric extinction.  We showed in
\S\ref{sec:colormag} and \S\ref{sec:atmext} that \slr\ corrects for
atmospheric attenuation in the $z$-band with sub-percent accuracy.
Our \slr\ formalism also supports color-airmass corrections, but we
have shown that we already achieve high quality atmospheric
corrections through a wide range of airmasses without these.

\paragraph{\slr\ corrects for attenuation through clouds.}

As long as the exposures are long enough to homogenize the extinction
due to clouds blowing across the images being analyzed, then the \slr\
technique will compensate for a common flux diminution across each
frame. Integration times that exceed 60 seconds typically satisfy
this, even for images that span a degree on the sky.
%FWH: I don't see how this is true:
% Even if the
% extinction due to clouds is not ``grey,'' we will obtain properly
% corrected colors. 
The multi-epoch \sdss\ Stripe 82 analysis of
\citet{bib:ivezic_stripe82} supports the assertion that we can treat
the effect of clouds as ``grey'' extinction, which \slr\ corrects for
at once with $\zptcolor$.

\paragraph{\slr\ corrects for passband sensitivity differences, even
  across different instruments and telescopes.}

An overall multiplicative difference between the system throughput
vs.\ wavelength of two different cameras or telescopes can be caused
by filter transmission functions, detector quantum efficiency, and
other instrumental properties.  These are some reasons the nominal
zeropoints between two facilities are not equal.  We have shown that
\slr\ accounts for zeropoints in all tests we performed
(\S\ref{sec:tests}).  {\it Differential} variation in system
throughput vs.\ wavelength brings about disagreements in apparent
magnitudes and colors that change with objects' color.  The \slr\
technique takes these into account by applying traditional color
terms, measured from infrequent standard star observations.  In
color-color space, these cause apparent colors to scale, shear, and
rotate.  Our \slr\ methodology fully permits color term correction,
and we have demonstrated this in \S\ref{sec:colormag} and implicitly
in \S\ref{sec:photoz} using two different cameras.

\paragraph{\slr\ circumvents aperture corrections and photometric
  artifacts from PSF variations.}

Because the \slr\ technique maps observed, instrumental magnitudes
onto a standard stellar locus, even if photometry from a given image
in a given passband has some systematic photometry error from
PSF-related issues, the applied color transformation will correct for
these artifacts.  Differences in aperture corrections between
passbands produce a common displacement in color-color space.
\citet{bib:ivezic_stripe82} showed that using the stellar locus to
correct colors of galaxies works well for photometry obtained with the
\sdss\ analysis pipeline. We have demonstrated in \S\ref{sec:photoz}
that applying \slr\ to Source Extractor instrumental magnitudes
produces photometric redshifts that are in excellent agreement with
spectroscopic redshifts, without additional aperture corrections.

\paragraph{\slr\ avoids the need to take more than a single
  calibration frame per instrument, per filter.}

The universality of the stellar locus in effect allows the colors from
a set of multiband images to be self-calibrating. We find that taking
only one set of multiband images of a calibration field is required,
to determine the instrumental color terms.  At all times thereafter,
on timescales of months or years, and as long as there is a sufficient
number of stars ($\gtrsim 10$) in each image being analyzed, there is
no need to obtain additional calibration frames.  Our galaxy cluster
observations of \S\ref{sec:photoz} used color terms measured only once
in an \sdss\ standard star field.

\paragraph{\slr\ corrects for Galactic extinction, independent of
  $A_V$ and $R_V$.}

Whatever processes might bring about an overall observed displacement
of the stellar locus, \slr\ will make the appropriate correction as
long as it is locally well correlated. This includes Galactic
reddening effects when the stars are behind the dust.  Because a
single global color correction is applied to all the data, the
cataloged objects should span a sufficiently small region on the sky
to assure they have a common Galactic reddening.  We showed in
\S\ref{sec:galext} that \slr\ can recover the canonical reddening
through very high and very low dust thicknesses to
$\sim20\,\mathrm{mmag}$ in color.  We point out that our photometric
redshift analysis in \S\ref{sec:photoz} made no explicit dust
correction of any kind, and results were nonetheless consistent with
an intrinsic Galactic dereddening by \slr.  The anomalous extinction
results of \S\ref{sec:colormag} are suggestive of possible deviations
from the typically assumed reddening law---although drifts in mean
metallicity cannot be ruled out as yet.

\paragraph{\slr\ takes advantage of all stars in the images, producing
  a ``democratic'' color calibration.}

For fields of view that are currently typical of astronomical imaging
instruments, Galactic stars are approximately uniformly distributed
across the image. \slr\ allows for the calibration of all objects in a
field using the very stars appearing in that field.  As long as there
are no effects that vary across the field of view, the \slr\ color
calibration technique inherently produces a homogenized calibration of
colors across the frame.

\paragraph{\slr\ uses a single common standard stellar locus, and
  thereby reduces sensitivity to systematic errors in photometric
  zeropoints.}

The photometry from \sdss\ exhibits zeropoint drifts at the few
percent level across the sky \citep{bib:ubercal}.  If these cataloged
magnitudes are used to derive local colors, the colors inherit a
position-dependent systematic drift that is the difference between the
underlying photometric zeropoint errors.  \slr\ uses a single common
standard stellar locus for all color corrections, so the derived
colors should show reduced position-dependent systematic errors.  The
{\it precision} of \slr-derived colors should be outstanding.  The
{\it accuracy} of \slr-derived colors depends upon the accuracy of the
standard stellar locus, the extent to which this locus is
representative of the population being observed, and the nature of the
Galactic dust extinction, as discussed in \S\ref{sec:stellarlocus}.

\paragraph{\slr\ is flexible.}

The simple matching of the observed instrumental stellar locus to a
standard one allows for a diversity of input catalogs. The input can
be purely flat-fielded, instrumental photometry, and \slr\ will
correct them for atmospheric and Galactic extinction, and zeropoints.
\slr\ will equally allow input photometry that has already been
corrected for some or all of these effects, as previous authors have
shown. We have applied \slr\ to pre-calibrated \sdss\ photometry
(\S\ref{sec:colormag}) and to instrumental photometry from two
different instruments (\S\ref{sec:atmext}, \S\ref{sec:photoz}).  The
method makes no distinction between the various sources of locus
shifts in color-color space, so it can be applied to photometric
catalogs generated up to any stage in the traditional analysis chain.
One can even envision mixing catalogs by, for example, acquiring
$griz$ imagery of a field and performing \slr, then acquiring further
$z$-band images of the same field and coupling the new instrumental
$z$-band catalogs to the already-calibrated $gri$ ones.  The resulting
$\zptcolor$ will produce the new $z$-band calibration, and should
reproduce the $gri$ calibrations.  \slr\ can be optimized to give fast
calibrations for a wide variety of optimized observing strategies. Of
course, \slr\ will perform best when the prior reduction steps do not
induce additional color scatter nor scalings, rotations, and shears.

\paragraph{\slr\ is fast: mountaintop reductions can provide accurate
  colors, and hence photometric redshifts, on-the-fly.}

Optimizing the use of allocated telescope time is an important
goal. By performing real-time \slr\ reductions of photometric data,
observers can determine when a desired signal to noise ratio is
attained. A specific application of \slr\ by our group is the
real-time determination of photometric redshifts of large numbers of
galaxy clusters---an application we illustrated in \S\ref{sec:photoz}.
In upcoming runs we intend to use \slr\ to make adaptive adjustments
of integration times while observing clusters of known position but
unknown redshift.

\subsection{Limitations}
\label{sec:limitations}

As a different way of calibrating photometry, \slr\ also carries some
unique limitations.  The main limitation is that \slr\ performs better
in obtaining highly accurate colors than magnitudes in each of the
bands.  By the same token, the colors that are produced by the \slr\
approach are only as good as the standard stellar locus used as the
calibration standard.  Of particular concern is the need of identify a
set of calibration stars that have suffered minimal Galactic
extinction, and whose median metallicity reflects that of the stellar
populations we observe in practice.

Another principal limitation is that \slr\ calibrations
necessarily correct for Galactic extinction.  To \slr, the dust
correction is as natural as the atmospheric extinction correction.
This is because our locus is standardized to stars suffering minimal
extinction, while stars we observe in practice will almost always be
behind the dust.  It is, however, more common to calibrate photometry
to the top of the atmosphere and 
% {\it in front of} the Galactic dust, and
only optionally apply an SFD Galactic extinction correction by hand, if needed.
There is considerable uncertainty in our understanding of the dust,
and this must be folded into our {\it a priori} color error estimates.
In nearly all the fields we have studied, results suggest that \slr\
corrects extinction through a wide range of dust thicknesses with high
accuracy.  Users should take special care not to double-correct the
extinction when using \slr.

Consequently, \slr\ will produce discrepant results if the sources of
extinction vary significantly across the field of view.  This
introduces a trade-off between larger field size, which allows more
stars to be included in the regression, and spatially constant
extinction, which \slr\ corrects for best.  We have applied \slr\ on
small fields of view with as few as $7$ useful stars, and the
uncertainties in these cases are predominantly statistical.  If \slr\
is to be applied to larger fields, one possible way to minimize extra
scatter from Galactic extinction is to first apply the SFD dust
correction for each stellar position, and then allow \slr\ to make the
residual color calibration.  For widely varying dust across a field,
this should suppress the large scale gradients and thus unnecessary
scatter in the stellar locus.

A third source of concern, both for traditional methods and \slr, are
systematic differences between point source and galaxy photometry.
\citet{bib:ivezic_stripe82} and our own results demonstrate that
applying a color correction derived from stars to galaxies does
produce reliable photometric redshift estimates for red sequence
galaxies.  Some level of systematic error is nonetheless expected,
because the spectral energy distributions (SEDs) of Galactic stars
differ from the underlying SEDs of possibly redshifted galaxies of all
types.  Moreover, heavily extincted sources with different intrinsic
SEDs will not suffer identical color shifts \citep[][]{bib:mccall}.
These are fundamental limitations of broadband photometry: as
integrals over the product of a source's SED, atmospheric and Galactic
dust transmission functions, and the instrumental sensitivity curve,
broadband measurements entangle the true SEDs with intervening
attenuation effects almost irreversibly \citep[][]{bib:endtoend}.
This systematic error afflicts {\it all} photometry that uses stellar
calibration standards, and is not unique to \slr.

%FWH: This goes to Section 2
% For the \slr\ implementation described here, a single common color
% correction is made to all objects in the photometric catalog. This
% assumes that all stars lie behind the extinction layers, namely the
% atmosphere and any sources of Galactic extinction. The first of these
% assumptions is certainly valid, since stars do indeed lie above the
% Earth's atmosphere.  The relative locations of objects and Galactic
% dust regions along the line of sight is more problematic, especially
% at low Galactic latitude.  \citet{bib:juric} have used the magnitudes
% and colors of the \sdss\ stars to roughly locate the source Galactic
% extinction, and they concluded that the overwhelming majority of
% \sdss\ stars lie behind the dust lanes in the Milky Way.
% \citet{bib:2mass_extinction} used the \tmass\ catalog to investigate
% the $3$-dimensional distribution of Galactic extinction.
% \citet{bib:IPHAS_extinction} have used H$\alpha$, $r$, and $i$
% photometry to map out $3$-dimensional extinction, particularly
% exploiting the fact that in their color-color space the Galactic
% extinction vector is not aligned with the stellar locus.  Colors of
% objects in the immediate solar neighborhood and colors of solar
% systems objects will {\it not} be treated properly with \slr.
% Similarly, any stars that suffer significant local extinction (for
% example in star forming regions) will distort the \slr\ analysis, and
% should be avoided. A sigma-clipped analysis can help suppress the
% systematic error from these outliers.

We finally note that, in the optical, \slr\ implicitly requires the
$g$-band.  This is because the critical optical feature that makes our
realization of \slr\ possible is the kink in the stellar locus at
$r-i\sim0.7$---the stellar locus in the $(r-i,i-z)$ plane is virtually
featureless.  The kink (1) gives the one-dimensional locus line a
distinctive shape that uniquely locates it in color-color space, and
(2) provides a component of the locus that is nearly perpendicular to
the reddening from Galactic and atmospheric extinction \citep[see
also][]{bib:baltic}.  A purely linear stellar locus line does not
allow for a {\it unique} stellar locus regression solution, as an
instrumental locus could slide freely along that line.  Inspection of
Figure \ref{fig:example} reveals the importance of having $g$-band
data.

% Finally, the \slr\ approach assumes that the observed stars have the
% same intrinsic distribution in color-color space as the stars in the
% standard stellar locus, to which we fit. A difference in the
% properties of the stellar populations, such as age and metallicity,
% could distort the \slr-derived colors---effects we explore in
% \S\ref{sec:stellarlocus}. We conservatively consider the present
% implementation of \slr\ to be reliable for Galactic latitudes
% $|b|>35\deg$.

\subsection{Future Directions}

\paragraph{Standard locus refinements.}

We consider the standard stellar locus used for this paper to be a
starting point.  \slr's performance rests on the standard locus being
representative of the population that any given observation probes.
It may be advantageous to develop different standard loci for
different expected populations---for example, one for low Galactic
latitudes, or different standard loci for different exposure times.
The various standard loci could be measured empirically, or the
expected perturbing effects could be modeled analytically and applied
to the \citet{bib:covey} locus.

\paragraph{A fuller treatment of Galactic extinction.}

As pointed out by \citet{bib:mccall}, extinction by Galactic dust
produces a shift in color that depends on the underlying photon
spectral energy distribution of the source. For the current
implementation of \slr\ we have ignored this effect, which will
produce shears, scalings, and rotations as well as a simple
translation in color-color space.  These are analogous to
color-airmass effects.  We also expect that, if large fields must be
calibrated in one \slr\ pass, it may be advantageous to apply the SFD
correction for each stellar position first, as mentioned in
\S\ref{sec:limitations}.  This may reduce scatter in stellar loci
constructed from large data sets with widely varying extinctions.

\paragraph{\slr\ vs.\ Galactic extinction and metallicity.}

This paper has shown that \slr\ accounts for the Galactic extinction
in nearly all fields we have studied so far.  The notable exceptions
are the Stripe 82 results in \S\ref{sec:colormag}, which corroborate
those of \citet{bib:ivezic_stripe82}.  Interestingly, the Galactic
latitudes of the Stripe 82 fields where stellar locus methods fail are
roughly equal to and {\it higher than} many of the fields we have used
to recover the SFD extinction (\S\ref{sec:galext}) and measure cluster
redshifts (\S\ref{sec:photoz}).  This presents difficulty in placing a
hard lower bound on $|b|$ where \slr\ should be valid.  While we
expect \slr\ to perform more reliably toward higher Galactic
latitudes, we suspect \slr, by its nature, has the potential to make
{\it better} Galactic extinction corrections than SFD.  This warrants
extended and in-depth studies comparing \slr\ color shifts jointly to
SFD prediction and metallicity.

\paragraph{Other passband combinations.}

We have concentrated heavily on the $grizJ$ bands, but the basic
approach should be applicable to other photometric systems, such as
$u$, $HK$, and $UBVRI$.  \citet{bib:odts1} have shown that the stellar
locus as probed by the Johnson bands has a kink feature that makes
this possible.  We will also be interested to see how the PanSTARRS
$y$ band at $1\,\mu\mathrm{m}$ will add to the accuracy of \slr. Another
interesting path to pursue is the use of the \slr\ approach to
transform data between photometric systems.

\paragraph{Multi-frame \slr.}

If we ignore color-airmass terms, then image stacks from individual
frames taken at different airmasses should also be amenable to \slr\
analysis.  A better approach, however, might be to extract \slr\
colors from all possible independent permutations of the multiband
images, and then average the resulting colors.  This is similar to the
philosophy applied in the ``$N(N-1)$'' approach to frame subtraction
photometry for supernova cosmology \citep{bib:NN1}, wherein all
possible pairs of images are subtracted.  This seems especially worth
pursuing for multi-epoch surveys like PanSTARRS and LSST, as it
completely sidesteps the challenges of combining frames with different
PSFs obtained at different airmasses.

\paragraph{Color recalibration of \sdss\ and \tmass?}

If the stellar locus does in fact provide us with a uniform
calibration source over most of the sky, and if \slr\ provides special
insensitivities to anomalous atmospheric and Galactic extinction
effects, then it is enticing to consider a joint recalibration of {\it
  colors only} from both the \sdss\ and \tmass\ surveys. A Bayesian
approach using the uniformity of the stellar locus, as was done in
optical bands by \citet{bib:ivezic_stripe82}, might improve the
accuracy of both optical and IR colors in these catalogs.

%%%%%%%%%%%%%%%%%%%%%%%%%%%%%%%%%%%%%%%%%%%%%%%%%%%%%%%%%%%%%%%%%%%%%%%%%%%%%%%%

\section{Conclusions}
\label{sec:conclusions}

We have developed and demonstrated a technique that exploits the
universality of the stellar locus to immediately obtain accurate
colors from uncalibrated multiband data.  For those who might wish to
exploit the \slr\ approach, the core IDL tools we have developed are
available at \url{http://stellar-locus-regression.googlecode.com}.

% \slr\ is inherently sensitive to Galactic extinction.  
Using archived
photometry from the SDSS survey, we have demonstrated that \slr\ can
produce results that agree with the commonly used SFD extinction map
over a wide range of dust thicknesses.  The performance of \slr\ in
these cases appears to be limited by systematic zeropoint drifts in
SDSS magnitudes.  We also reproduced the anomalous stellar locus
reddening results of \citet{bib:ivezic_stripe82} in a few \sdss\
Stripe 82 fields.  This puts into question (1) the validity of the
$R_V=3.1$ reddening law in those fields, (2) whether the dust is {\it
  both} in front of and behind the stars, and (3) whether these fields
have stars with spatially correlated deviations from expected median
metallicity.  

Images of fields obtained through a wide range of airmasses were subjected to \slr\ analysis.  We
recovered the colors of the stars in these images with an uncertainty
limited by the Poisson and flat-field errors in the photometry from
each frame.  Using \slr-only techniques and \tmass\ calibrated
photometry, we additionally obtained $i$-band zeropoints in the fields
good within $18\,\mathrm{mmag}$.

Finally, we have also presented photometric redshift results using
\slr-derived colors only.  We recovered the spectroscopic redshifts of
$11$ galaxy clusters at redshifts $0.09<z<0.25$ with
$\sigma_z/(1+z)=0.6\%$ RMS residual error.  The redshift residuals
also showed that the red sequence galaxy colors from \slr\ alone were
consistent with an intrinsic Galactic extinction correction.

The \slr\ technique, as implemented in our IDL code, can be used at
the telescope in real-time to optimize the use of allocated telescope
time.  We regard \slr\ as a promising way to calibrate colors and
magnitudes using fundamentally different physical assumptions,
providing calibrations far faster than the traditional approach.

%%%%%%%%%%%%%%%%%%%%%%%%%%%%%%%%%%%%%%%%%%%%%%%%%%%%%%%%%%%%%%%%%%%%%%%%%%%%%%%%

\acknowledgments

The approach outlined here builds upon the impressive work carried out
in building and analyzing the \sdss\ and \tmass\ data archives. We are
grateful to the builders and operators of the \sdss\ and \tmass\
systems.  We also acknowledge our deep reliance on the groundwork laid
by the authors of the \citet{bib:covey},
\citet{bib:ivezic_metallicity}, \citet{bib:sdss_variables}, and
\citet{bib:ivezic_stripe82} papers.  We are also grateful to the
developers \citep{bib:isochrones} of the on-line stellar evolution
modeling tool at {\tt http://stev.oapd.inaf.it/cgi-bin/cmd\_2.1}. We
are pleased to thank D.\ Finkbeiner, Z.\ Ivezic, T.\ Axelrod, A.\
Saha, D.\ Burke, J.\ Mohr, C.\ Smith, W.\ M.\ Wood-Vasey, A.\ Loehr,
T.\ Stark, N.\ Suntzeff, J.\ Tonry, and J.\ Battat for valuable conversations.
 
This research has made use of the NASA/IPAC Extragalactic Database
(NED) which is operated by the Jet Propulsion Laboratory, California
Institute of Technology, under contract with the National Aeronautics
and Space Administration.  This publication has made use of data
products from the Two Micron All Sky Survey, which is a joint project
of the University of Massachusetts and the Infrared Processing and
Analysis Center/California Institute of Technology, funded by the
National Aeronautics and Space Administration and the National Science
Foundation.  This research has made use of the NASA/ IPAC Infrared
Science Archive, which is operated by the Jet Propulsion Laboratory,
California Institute of Technology, under contract with the National
Aeronautics and Space Administration.

Funding for the creation and distribution of the SDSS Archive has been
provided by the Alfred P. Sloan Foundation, the Participating
Institutions, the National Aeronautics and Space Administration, the
National Science Foundation, the U.S. Department of Energy, the
Japanese Monbukagakusho, and the Max Planck Society. The SDSS Web site
is http://www.sdss.org/.  The SDSS is managed by the Astrophysical
Research Consortium (ARC) for the Participating Institutions. The
Participating Institutions are The University of Chicago, Fermilab,
the Institute for Advanced Study, the Japan Participation Group, The
Johns Hopkins University, Los Alamos National Laboratory, the
Max-Planck-Institute for Astronomy (MPIA), the Max-Planck-Institute
for Astrophysics (MPA), New Mexico State University, Princeton
University, the United States Naval Observatory, and the University of
Washington.

This work is supported by the NSF (AST-0607485), the DOE
(DE-FG02-08ER41569), and Harvard University. We also thank the team of
scientists, engineers and observing staff of the Las Campanas
Observatory.

{\it Facilities:} \facility{Magellan:Baade (\imacs)},
\facility{Magellan:Clay (\ldss)}, \facility{CTIO:2MASS},
\facility{FLWO:2MASS}, \facility{Sloan}

%%%%%%%%%%%%%%%%%%%%%%%%%%%%%%%%%%%%%%%%%%%%%%%%%%%%%%%%%%%%%%%%%%%%%%%%%%%%%%%%

\appendix

\section{Motivating the Color Transformation Equation}
\label{app:corresp}

Equation \eqref{eqn:transform} represents photometric calibrations using a compact, filter-independent matrix notation because
in the future we wish our procedure to be applicable to any set of
filters.  In this appendix we give a concrete example for the $griz$
SDSS filters in order to motivate and convey the meaning of the notation.

\subsection{Usual Photometric Calibration Equations}

Photometric calibrations are normally modeled with simple additive and
sometimes multiplicative terms.  Following the southern SDSS standards
literature \citep{bib:smith02,bib:smith03,bib:clem}\footnote{For other
  relevant unpublished documents see
  http://www-star.fnal.gov/Southern\_ugriz/publications.html} as rough
guides, we relate the instrumental magnitude $(g,r,i,z)$ through
the SDSS passbands to the true, extra-Galactic magnitude
$(g_0,r_0,i_0,z_0)$ as
\begin{subequations} 
  \label{eqn:photometry}
\begin{align}
  g & = g_0 + a_g + E_g + A_g + b_g(g_0-r_0) + c_gX_g(g_0-r_0) \\
  r & = r_0 + a_r + E_r + A_r + b_r(r_0-i_0) + c_rX_r(r_0-i_0) \\
  i & = i_0 + a_i + E_i + A_i + b_i(i_0-z_0) + c_iX_i(i_0-z_0) \\
  z & = z_0 + a_z + E_z + A_z + b_z(i_0-z_0) + c_zX_z(i_0-z_0) \\
  J & = J_0,
\end{align} 
\end{subequations} 
where in the last equality we take the \tmass\ data to be calibrated
already.  Here, $a_{n}$ is the zeropoint for the passband $n$; $E_n$
is the atmospheric extinction, often modeled as $k_nX_n$ for some
filter-dependent coefficient $k_n$ and airmass $X_n$ through which
exposure $R$ is taken; and $A_n$ is the estimated Galactic extinction.
$b_n$ is the color coefficient, and $c_n$ is the color-airmass cross
term coefficient.  The free parameters $k_{n}$, $a_{n}$, $b_{n}$ and
$c_{n}$ can be estimated from the literature or measured using
intermittent standard star observations, interpolating in airmass and
in time to the science exposures.  Galactic extinction is estimated
using SFD.

\subsection{Corresponding Color Equations}

Colors are magnitude differences.  Subtracting magnitudes between
adjacent passbands using Equations \eqref{eqn:photometry} gives
\begin{subequations} 
\label{eqn:color}
\begin{align}
  (g-r) & = (g_0-r_0) + (a_g-a_r) + (E_g-E_r) + (A_g-A_r) + \\
  & b_g(g_0-r_0) - b_r(r_0-i_0) + c_gX_g(g_0-r_0) - c_rX_r(r_0-i_0) \\
  (r-i) & = (r_0-i_0) + (a_r-a_i) + (E_r-E_i) + (A_r-A_i) + \\
  & b_r(r_0-i_0) - b_i(i_0-z_0) + c_rX_r(r_0-i_0) - c_iX_i(i_0-z_0) \\
  (i-z) & = (i_0-z_0) + (a_i-a_z) + (E_i-E_z) + (A_i-A_z) + \\
  & b_i(i_0-z_0) - b_z(i_0-z_0) + c_iX_i(i_0-z_0) - c_zX_z(i_0-z_0) \\
  (z-J) & = (z_0-J_0) + a_z + E_z + A_z + \\
  & b_z(i_0-z_0) + c_zX_z(i_0-z_0).
\end{align}
\end{subequations} 
This is greatly simplified with matrix notation.  We define the
instrumental color vector as
\begin{equation}
\color \equiv
\begin{pmatrix}
g-r\\r-i\\i-z\\z-J
\end{pmatrix},
\end{equation}
and likewise for the true color vector $\color_0$; the additive color
calibration vector as
\begin{equation}
\zptcolor \equiv
\begin{pmatrix}
a_g-a_r+E_g-E_r+A_g-A_r \\
a_r-a_i+E_r-E_i+A_r-A_i \\
a_i-a_z+E_i-E_z+A_i-A_z \\
a_z+E_z+A_z \\
\end{pmatrix};
\end{equation}
the color term matrix as
\begin{equation}
\colormatrix \equiv
\begin{pmatrix}
b_g & -b_r   & 0         & 0  \\
0     & b_r  & -b_i      & 0 \\
0     & 0      & b_i-b_z & 0 \\
0     & 0      &     b_z & 0 \\
\end{pmatrix};
\end{equation}
and the color-airmass term matrix as
\begin{equation}
  \colorairmassmatrix \equiv
\begin{pmatrix}
c_gX_g & -c_rX_r   & 0         & 0  \\
0     & c_rX_r  & -c_iX_i      & 0  \\
0     & 0      & c_iX_i-c_zX_z & 0  \\
0     & 0      &        c_zX_z & 0  \\
\end{pmatrix};
\end{equation}
then we can write the color calibration equations as a single matrix
equation,
\begin{equation}
  \label{eqn:color_tensor}
  \color = \zptcolor + \left(\identity + \colormatrix +
    \colorairmassmatrix\right)\color_0,
\end{equation}
where $\identity$ is the identity matrix.  Setting
$\colorairmassmatrix=\boldsymbol{0}$, this is equal to Equation
\eqref{eqn:transform}.  The inverse color transformation is
\begin{equation}
  \label{eqn:invcolortransform}
  \color_0 = \left(\identity + \colormatrix +
    \colorairmassmatrix\right)^{-1}\left(\color- \zptcolor\right).
\end{equation}
Whether the transformation matrices are invertible depends on the
details of how first order color coefficients are chosen, but under
normal circumstances it will be possible.  Using the example we've
just presented, $(\identity+\colormatrix)$ is invertible because its
determinant is $(1+b_g+b_r+b_gb_r)(1+b_i-b_z)$, which is not equal to
zero for typical (small) values of color terms.

This is how our matrix notation corresponds to typical photometric
calibrations.  The notation is trivially generalized to any filter
set.  The notation also motivates generalizations in choice of
calibration parameters.  For example, the color term matrix
$\colormatrix$ can be made to be diagonal, (anti-)symmetric, or
populated entirely by different nonzero entries, depending only the
particular application.  Furthermore, higher order terms can be
considered by including higher powers of $\color_0$ to Equation
\eqref{eqn:color_tensor}.

\bibliography{ms}
%\bibliography{ms.bib}
%\bibliography{ms.bbl}

\end{document}